\documentstyle[prl,aps,preprint]{revtex}
\begin{document}

\begin{center}
   {\Large \bf  Dynamics of particle deposition on a disordered substrate:
   II. Far-from Equilibrium behavior.
}
\end{center}
\begin{center}
Yan-Chr Tsai $\footnote[2]{Present address: 
Department of Physics and Astronomy,
 University of Pennsylvania,  Philadephia,  
 PA  19104-6396}$ and Yonathan Shapir
 \vskip 0.25in
Department of Physics and Astronomy,
University of Rochester, Rochester, NY 14627

\end{center}

\begin{abstract}

       The deposition  dynamics   of particles (or the growth of a  rigid
crystal) on  a disordered substrate at a finite deposition rate is explored.
We begin with an equation of motion which includes, in  addition to
the disorder, the periodic potential due to the discrete size of 
the particles
(or to the lattice structure of the crystal) as well as the term introduced
by Kardar, Parisi, and Zhang (KPZ)   to account for the lateral growth
 at a finite 
growth rate.  A generating functional for the 
correlation and response functions of 
this process is derived using the approach of Martin, Sigga, and Rose. 
A consistent renormalized perturbation expansion to first order in the 
non-Gaussian
couplings  requires the calculation of  diagrams up to three loops.
To this order we show, for the first time for  
this class of models which violates the 
the fluctuation-dissipation theorem,  that the theory is renormalizable.

We find that the effects of the periodic potential and the 
disorder decay on very large
scales and  asymptotically the KPZ term dominates the behavior. 
However, strong non-trivial
crossover effects are found for large intermediate scales.
\end{abstract}

\section{Introduction}
1.\underline {General.}

A few years ago the near equilibrium dynamics of a growing 
crystalline surface
was elucidated \cite{Week,Growth}. It was found that a 
roughening transition occurs at $T=T_{r}$
between a high-temperature rough phase and a phase with a 
flat surface for
$T<T_{r}$. The mobility of the growing surface drops from a 
finite value to
zero as $T\rightarrow T_{r}^{+}$. In the low-temperature 
phase the growth is "activated"
with the formation of higher "islands" on top of the 
flat surface. Similar 
behavior occurs in deposition of cubic particles with 
diffusion on a flat substrate. The transition
in this case is as function of the inherent noise due to spatial and
temporal fluctuations  in the
deposition.

In view of the existence of a low temperature (or low noise) 
flat phase, the question
of how disorder in the substrate will change the behavior had to 
be addressed.
We have initiated a comprehensive study of the related questions. 
A short letter which
announces the novel and surprising results was published 
elsewhere \cite{TS}. 
In a previous full paper \cite{TS1} (denoted by I in the
following) we have presented the detailed calculation and analysis 
for the near-equilibrium
dynamics. In this regime the averaged growth rate is small. 
The equation of motion is 
derived from the Hamiltonian of the system. Detailed-balance 
and the Fluctuation-Dissipation
Theorem (FDT) \cite{SMS} both hold. Yet we found very non-trivial 
results: super-rough correlations,
temperature-dependent dynamics exponent, and a non-linear 
relation between the average
growth rate and the driving force, were found below a 
super-roughening
transition temperature $T_{sr}$.

In the present paper, the second in the series, we present 
our detailed calculations and 
results for the  dynamics far-from  equilibrium. In this case 
the equation-of-motion
cannot be derived from a Hamiltonian. Detailed-balance and 
the FDT are both violated.
That situation represents a much more serious theoretical 
challenge since even the very 
renormalizability of the  process is questionable.

The equation of motion we analyze describes the deposition 
of cubic particles
on a random substrate. It will also apply to the surface 
of a crystal if the solid is
very rigid. Since the effects  of the disorder in the 
substrate will be felt only up to
 a height
$h^{*}$ (which is larger if the solid is more rigid), our 
theory will apply as long as 
$h<h^{*}$.

The  width of  a growing surface $w$ follows generally 
the scaling form \cite{GR1,krug}: 
\begin{equation}
w(L,t) \sim L^{\alpha} f(t/L^{z}), 
\end{equation} 
where  $t$ is the time and $L$ is the linear size of the system.
$\alpha$ is the roughening exponent and $z$ is the  dynamic exponent.

The leading difference between near- and far-from equilibrium 
is due to the lateral 
growth. If the growth rate is finite the lateral growth add 
a term proportional to
$(\nabla h)^2$ to the equation of motion. This term was derived 
by Kardar, Parisi
and Zhang (KPZ) \cite {KPZ} as the most relevant term in the 
expansion in term of $\nabla h$.

As   the RG analysis shows, the KPZ nonlinearity is  
marginally  relevant  in 2+1 
dimensions.  The asymptotic behavior  is controlled by this 
nonlinearity,
which violates the FDT.    Consequently, the roughness
 of the growing surface is also
affected by the nonlinearity.  Several  simulations showed 
 $\alpha \sim  0.4$ [9,10]
which is larger  than that of the near-equilibrium case ($\alpha =0$).
Many variants of KPZ-related models have been studied
in recent years [6,7,11-22]. 

The effects of such a term on the growing crystalline surface 
on a flat substrate was
considered by  Hwa, Kardar, and Paczuski (HKP) \cite{HKP}. 
We review and revise their results in the
next section of this Introduction (Ch.I). Ch. II will be 
devoted to the derivation
of the  generating functional, and the description of 
the RG procedure.
Ch. III, IV and V are devoted to the outline of the 
calculations of the different renormalization
factors. Extensive details of these calculation are 
relegated to the appendices. 
In Ch. VI the recursion relations are derived. Ch. VII 
is devoted to the analysis of the asymptotic
and crossover behaviors  which follows from  these recursion 
relations.  
Our main conclusions are summarized in the last chapter VIII.

2.\underline {Review  of  previous works}

The system we shall study in this paper has three 
important and non-trivial ingredients:

  (i) The periodic potential,
  (ii) The disorder in the substrate, and 
  (iii) The KPZ non-linearity arising from the lateral growth 
in the presence of a finite 
      driving force (or deposition rate).

Paper (I) discussed the case in which (i)+(ii) are present. 
The major physical
consequences, namely the existence of a super-rough 
"glassy phase" for $T<T_{sr}$ with 
intriguing static
and dynamic properties, were  discussed there and will not 
be elaborated further here.
Our goal here is to see how the addition of (iii) (i.e. the KPZ term) 
modifies the behavior.

However, we can also take another point of view and 
ask how the addition of (ii) (i.e.
the disorder in the substrate) is modifying the behavior 
found in presence of (i)+(iii).
The model which discussed the non-equilibrium growth on a 
flat surface in presence of 
both the periodic potential and the KPZ non-linearity was 
analyzed by HKP \cite{HKP}.
The equation of motion which describes the growing surface 
under these calculation
is: 
\begin{equation}
\tilde {\mu} ^{-1}\frac {\partial h(\vec {x},t)}{\partial
t}=F+\nu(\nabla ^2
h(\vec {x},t))+ \frac {\lambda }{2} (\nabla h)^2 + \frac {\gamma
y_2}{a^2}\sin
[\gamma (h(\vec {x},t))] +\zeta (\vec {x},t),
\label{eq:mot}
\end{equation}

In this equation $h(\vec x, t)$ is the local height at 
time $t$, $\vec x= (x,y)$ are the
coordinates in the 2d basal plane; $\mu$ is the 
microscopic mobility; $F$ is the driving
force; $\nu$ is the surface tension; $\lambda  $ is 
the coefficient of the KPZ non-linearity
(which is proportional to $F$); $y_2$ is the  coefficient 
of the leading harmonies 
(higher harmonies are 
irrelevant and the lattice spacing is taken to be 
unity for simplicity);
$\gamma=\frac {2\pi}{b}$, where $b$ is the vertical 
lattice spacing ;$a$ is the horizontal lattice spacing; and
 $\eta (\vec x.t)$
is the noise in the deposition (or due to thermal fluctuations) 
which obeys:
\begin{equation}
<\eta (\vec x , t)\eta (\vec x^{'}, t^{'})>=2D \delta(\vec x 
-\vec x ^{'}) \delta (t-t^{'})
\end{equation}

They obtained a set of recursion  relations from which 
they reached several conclusions.
They found a critical temperature $T_{c}$. For $T>T_{c}$, 
$y_2$ decays slowly to zero 
and the large scale behavior is determined by the KPZ 
coupling. Approaching $T_c$
from above the linear-response macroscopic mobility 
(Namely the ratio $v/F$ where
$v=<\frac{d h}{dt}>$ is the averaged growth rate in 
the limit $F \rightarrow 0$) vanishes as $(\ln |T-T_{c}|)^{-\eta }$.

In the low temperature  phase $T<T_c$ they have found that 
$y_2$ grows indefinitely large and therefore
they concluded that the surface is flat. 

While reanalyzing their work we have discovered a term 
that was overlooked in their
calculations and which might affect their latter 
conclusion. Indeed it may be shown 
that a term of the form $y_1 \cos(2\pi h)$ is generated 
under renormalization 
from the contraction  of the terms $\lambda  
(\nabla h)^2 $ and $y_2 \sin(2\pi h)$. 
This term also feeds back into the renormalization of $y_2$.

The recursion relations to lowest order are:
\begin{eqnarray}
\frac{dy_1}{dl}&=&(2-\frac{\pi \mu D}{\nu})y_1 -\rho 
\frac{\lambda }{\nu} y_2 ,\\
\frac{dy_2}{dl}&=&(2-\frac{\pi \mu D}{\nu})y_2 +\rho 
\frac{\lambda }{\nu} y_1,
\end{eqnarray}
where $\rho =\frac{1}{2}[4\pi ^2 \ln (4/3)(\frac{\mu D}{\nu})^2+
(\frac{\mu D}{\nu})] $.
The two harmonic terms may be combined into a single term:
\begin{equation}
|y| \sin [ 2\pi h +\vartheta (l)]
\end{equation}
with $y^2=y_1 ^2 + y_2 ^2$, and $\vartheta (l)=tg ^{-1} y_1/y_2$.

If we look at the flow of $|y|$ and $\vartheta$ we find 
that indeed $|y| 
\rightarrow  \infty$ for 
$T<T_c$ which would indicate a flat surface if $\vartheta$ 
was to remain constant.
However, the recursion relations   imply that the phase 
shift angle is rotating 
with $l$ like: $\vartheta (l)=w l$ with an  angular velocity 
$w \sim \lambda  /\nu$.
Since $l=\ln \tilde {b}$ where $\tilde{b}$ is the 
rescaling factor
that means an ever changing  $\vartheta \sim w \ln \tilde {b}$, 
with rescaling.

Therefore it  is not obvious that the surface is  flat. 
Hopefully, higher-order terms 
in the recursion relations of $y_1, y_2$ and $\nu$ will 
help to identify with more 
confidence the nature of low-temperature phase.

3.\underline {Introducing the disorder}:

As we explained in I, and as is clear from Fig. 1, the 
disorder will shift the origin of
the  periodic potential by a random and uncorrelated amount 
at every point $\vec x$
in the basal plane \cite{TS,TS1}.

We take the deposited 
 particle to have a rectangular shape with a square base 
of linear extent $a$
and height $b$ in the growth direction.
Then  the periodic potential becomes proportional to 
$\sin [\frac{2\pi}{b}(h(\vec x,t)+
d(\vec x))]$, where $d(\vec x)$  are the local deviation  in 
the height of the substrate. 
Let  us denote the associate phase $\Theta (\vec x)= 2\pi d(\vec x)/b$.
It obeys:
\begin{equation}
<e^{-\Theta (\vec x)} e^{-\Theta (\vec y)}> =a^2 \delta ^2 
(\vec x -\vec y).
\end{equation}

We have assumed that $d(\vec x)$ are typically of order 
$b$ (or larger) and that if
correlations exist in $d(\vec x)$ at different $\vec  x$ 
they are at most short range 
 (in which case they fall in the same universality
class as the $\delta$ correlated disorder we study here)
The equation of motion we need to study is therefore:

\begin{equation}
\tilde {\mu} ^{-1}\frac {\partial h(\vec {x},t)}{\partial
t}=F+\nu(\nabla ^2
h(\vec {x},t))+ \frac {\lambda }{2} (\nabla h)^2 + \frac {\gamma
y}{a^2}\sin
[\gamma h(\vec {x},t)+\Theta(x)] +\zeta (\vec {x},t).
\label{eq:mott}
\end{equation}

\section {Generating Functional and Basic Diagrams.}
\label{sec 63}

The Martin, Sigga, and Rose (MSR) \cite {MSR,Zinn,janssen} 
method is utilized to obtain the generating
functional  for the correlation and response theory. An auxiliary 
field $i\tilde{h}(\vec x,t)$ 
is introduced to enforce the equation of motion through 
an integral representation of 
the  delta function.
Then the thermal noise and the disorder are averaged 
upon to yield the following generating functional:
\begin{eqnarray}
\langle Z_{\Theta }[\tilde {J},J] \rangle _{disorder}&=&\int\!
{\cal
D}\tilde {h }{\cal D}h  \exp\{\!
\int \! d^2 x dt [\tilde {D}\tilde {\mu} ^2 \tilde {h }^2 -\tilde {h
}(\frac
{\partial }{\partial  t}h -\tilde {\mu} \nu
\nabla ^2 h -\tilde {\mu}\frac {\lambda }{2} (\nabla h)^2 )] \nonumber \\
&&+\frac {\tilde {\mu} ^2 \gamma ^2 \tilde {g}}{2a^2}\int\! \int \!
d^2x dt dt'\, \tilde
{h}(\vec {x},t)
\tilde {h }(\vec {x},t')  \cos (\gamma (h (\vec {x},t)-h
(\vec {x},t'))\}.\nonumber  \\
\label{eq:dgen}
\end{eqnarray}

a.\underline {The Basic diagrams:}

As explained in details in I the Gaussian (quadratic) part of 
the "action" gives rise to
the bare response function 
\begin{equation}
\langle \phi (\vec {q},\omega )\tilde {\phi}(-\vec
{q},-\omega)\rangle =\frac {1}{\mu (q^2+m^2)+i\omega },
\label{eq:p1}
\end{equation}
which is depicted in Fig. 2.

The bare correlation function is :
\begin{equation}
\langle \phi (\vec {q},\omega )\phi (-\vec {q},-\omega )\rangle
=\frac {2D\mu ^2}{[\mu (q^2+m^2)]^2+\omega ^2},
\label{eq:p2}
\end{equation}
where $m$,  the mass of the field $\phi$,  is  introduced to  control    
the infrared divergences  in  the Feynman integrals.
(In the  two-dimensional  regularization, it
is notorious \cite {Zinn} that their  infrared and 
ultra-violet divergences 
will mingle together   without introducing masses for the fields.)

In the momentum and time representation, they are given by 
\cite{janssen}:
\begin{equation}
\langle \phi (\vec {q},t) \tilde {\phi }(-\vec {q},t')\rangle
=\theta (t-t') e^{-\mu (q^2+m^2)(t-t')},
\label{eq:p3}
\end{equation}

\begin{equation}
\langle \phi (\vec {q},t)\phi (-\vec {q},t')\rangle =\frac {D\mu
}{q^2+m^2}e^{-\mu (q^2+m^2)|t-t'|},
\label{eq:p4}
\end{equation}

where $\theta(t)=1$ for $t>0$ and $\theta(t)=0$ for $t<0$.
The bare vertex $\phi \phi ^{'} \cos (\phi -\phi ^{'})$
is drawn  in Fig. 3.

b.\underline { The renormalization group procedure:}

We follow the minimal subtraction  scheme. The renormalization
parameters are related to the bare ones by the so-called 
$Z$ factors. Minimal scheme consists
in extracting from the diagrams only the divergent parts   which 
are all expressed in terms of functions of the 
dimensionality of the system.

The bare and
renormalized vertex functions can be related by factors of
$Z_{\phi }$, $Z_{\tilde
{\phi}}$.
For instance,
\begin{equation}
\Gamma_{N,L}^{R}(q,\omega; \varsigma _R,m_R,\kappa)=
(Z_{\tilde {\phi }})^{\frac {N}{2}}
(Z_{\phi })^{\frac {L}{2}}
\Gamma _{N,L} (q,\omega;\varsigma_0,m_0,a),
\label{eq:rb}
\end{equation}
where $\varsigma _R$ and $\varsigma _0$ label  renormalized parameters
($g,\mu,\cdots$) and bare
parameters ($g_0, \mu _0,\cdots$), respectively.
$q$ and $\omega$ are the external  momentum and  frequency, 
respectively.
In the corresponding vertex function, $a$ is a short-distance
cutoff, and $\kappa$ is a mass scale.
Here $\Gamma_{
N,L}$ stands for the vertex function with $L$ external $\phi$ lines
and $N$ external $\tilde {\phi}$ lines. The factors, $Z_{\phi}$ and
$Z_{\tilde {\phi}}$, are set to remove the divergent parts of
the vertex function $\Gamma$.

The following renormalization constants are defined through the 
relations between
the bare and the renormalized  couplings \cite{Zinn,SGT,GS,amit}
\begin{equation}
D_{0}=Z_{D}D,\quad
g_{0}=Z_{g}g , \quad \lambda  _0 = Z_{\lambda } \lambda 
,\label{eq:rgs1}
\end{equation}
\begin{equation}
m_{0}^{2}\phi ^2 =m^2 \phi _{R}^2 ,\quad
\gamma_{0}^{2}\phi ^{2}=\gamma^2 \phi _{R}^2,\quad
\phi^2 =Z_{\phi }\phi _{R}^2,\quad
\tilde {\phi }^2 =(\tilde {Z}_{\tilde {\phi }}) \tilde {\phi
}_{R}^2,\quad
\label{eq:rgs2}
\end{equation}
where $\tilde{Z}_{\tilde{\phi}}= Z_{\tilde{\phi}}^2$.

In the next chapter we concentrate on the procedures to calculate 
the $Z$ factors. The details are relegated to the appendices.

The renormalized perturbation theory even to lower non-trivial
orders in $g$ and $\lambda $ has to be consistent order by order in
$\gamma$. It also requires the calculations of Feynman 
diagrams \cite{Zinn,amit}
with up to three loops.

\section {Calculations of $Z_D $ and $Z_\mu$}
\label{sec63}
The renormalization of $\mu$ is not  affected in the presence of
the KPZ nonlinearity since the associated vertex function
$\Gamma
_{1,1}$  comes with one external $\tilde
{\phi}$ and one external $\phi$ 
and the  basic vertex  $\lambda $ contains derivatives on its two
$\phi$ legs. Thus the factor $Z_{\mu}$
remains the same  as in paper I (equilibrium
dynamics), and so does the recursion relation for $\mu$.
On the other hand, the parameter $D$  suffers additional
renormalization of order $\lambda  ^2$. Basically, the renormalization of
$D$ from $\lambda  ^2$ is the same as that of $D$ encountered in 
the KPZ model.
Here, we still  focus on the same  vertex function
$\Gamma _{2,0}$  as we did in  the previous paper (I).

Obviously, the first nontrivial contribution  begins from the second
order in $\lambda $.  As shown in Fig. 5, the vertex function is 
modified by the associated integral with $\lambda  ^2$.
The corresponding integral is given by:
\begin{eqnarray}
&&(\frac {\mu _0 \lambda  _0 }{2})^2\frac {1}{(2\pi)^2}
\int _{-\infty}^{\infty}d^2\vec p
\int _{-\infty}^{\infty}    d\Omega \frac{1}{2\pi}
\frac {(2D_0\mu_0 ^2)^2p^2 \cdot p^2}
{[\mu _0 ^2(p^2+m^2)^2+\Omega ^2][\mu _0 ^2(p^2+m^2)^2+\Omega ^2]}
\nonumber \\
&&= \frac{1}{4}
\lambda  _0 ^2 (\mu _0 ^3 D_0 ^2)\int _{-\infty}^{\infty}    d^2 \vec p
\frac {1}{(2\pi)^2}
\frac {1}{(p^2+m^2)} \nonumber \\
&&= \frac{1}{4} \lambda _0 ^2 (D_0 ^2 \mu_0 ^3)[-\frac {1}{4\pi}
\ln (c m^2 a^2)].
\end{eqnarray}
As illustrated in Fig.5, 
the symmetry factor for this diagram is
2 and another factor 2 arises from the differentiation of the
external legs.
One more factor $\frac {1}{2}$ is  due
to the  coefficient from the expansion of interaction with power
$2$ for the utilization of  $\lambda  ^2$. With the combination 
of  Eq.~(D6)
in  paper I, we have:
\begin{equation}
-2D \mu ^2 =( Z _{\tilde {\phi}}) ^2 [-2 D_0 \mu_0 ^2
-\frac {D_0^2\mu _0 ^3}{2}\lambda  ^2 (-\frac {1}{4\pi}\ln 
(cm^2a^2))+\gamma ^2
\sqrt {c} g \mu _0 \ln (c m^2 a^2)] .
\end{equation}
Therefore
\begin{equation}
Z_D=1+\frac {D \mu \lambda  ^2}{4} (\frac {1}{4\pi}
\ln (c m^2 a^2))+\frac {\gamma  ^2 \sqrt {c}g}{2 D\mu}\ln (cm^2a^2).
\end{equation}

\section { The Calculation of $Z_{\lowercase{g}}$}
\label{sec64}

As in the harmonic model of paper (I), we
consider the vertex function $\Gamma _{2,0}(\vec q,t;-\vec q
,t^{'}) $ in the limit $|t-t^{'}|\rightarrow  \infty$ \cite{GS}.
The calculation of $Z_g$ will be   based on 
 Eq. (8.3)in GS \cite{GS}, where  the  renormalization  factor is
substituted  in that equation to make $\Gamma _{2,0}(\vec q,t;-\vec q
,t^{'}) $ ($\vec q \rightarrow 0$) finite.  Therefore to find out $Z_g$ 
is just to calculate the  renormalization of   $\Gamma_{2,0}$.
  
The contributions to $Z_g$  we need to sum are  of order
$g^2$ (as in paper I)  and  of order
$\lambda  ^2 g$.

The combination of $g$ and $\lambda  ^2$  leads to  two
types  of diagrams, 2-loop and 3-loop diagrams.
Some of the associated diagrams are canceled by each other
as shown in Fig. 6 and Fig. 7.
The other non-vanishing diagrams, including
six 2-loop and two 3-loop diagrams, are shown in Fig. 8 -- 
Fig. 15.
The  detailed calculations of 2-loop  integrals are given in
Appendix A, where we also  explain the   cancellation
of sub-divergences of some diagrams.  The sum of
the leading and  sub-leading   divergences  contributing  
to  $Z_g$   are listed in Eq.~\ref{eq:ggz} (see the third and fourth
terms).  

In the  Appendix B,  we present the detailed calculation of the leading
divergences in the 3-loop diagrams and also show that
they do not
contain any sub-divergence.

Now, what  remains  to complete the  3-loop results is just to
sum up the leading divergent terms in $\ln (c m^2a^2)$, which
essentially contribute to the recursion relations.
The $\ln (cm^2a^2)$ contribution of  the diagram in Fig. 14 will be
\begin{eqnarray}
\lefteqn{-\frac {1}{2}[(3.1-1)-2(3.1-2)-2(3.1-3)] }\nonumber  \\
&=& \frac {3}{2} I_B +2I_D +4I_E +20\frac {1}{2}\ln (\frac
{3}{4})-6 (\ln (\frac {4}{3}))^2 -4 \ln (2) -\frac {5}{2}
(\ln 2)^2 +22 \ln (\frac {3}{2}) \nonumber \\
 &&+ 3\Phi (1,2) -3\Phi (\frac {1}{2},2)-\Xi (\frac {1}{4},2)
\end{eqnarray}
The contribution from the diagram in Fig. 15 is:
\begin{eqnarray}
\lefteqn{ [-(3.2-1)+2(3.2-2)+2(3.2-3)] }\nonumber  \\
&=&\frac {1}{-\frac {3\epsilon  }{2}}[\frac {1}{4}I_C +I_D-2I_E
+\frac {1}{2}\ln (\frac {3}{2})-\frac {3}{4}\ln (\frac {4}{3})
-\frac {1}{2}\ln (\frac {4}{3})].
\end{eqnarray}                                    

Now we are in a position to calculate  $Z_g$.
Inserting the above calculation results into the self energy
in reference \cite{GS}, we obtain:
\begin{eqnarray}
Z_g &=&1-\delta  _0 \ln (c m^2 a^2) +\frac {\lambda  ^2
\gamma ^2 (D\mu)^2}{8}\frac {[\ln (cm^2a^2)]^2}{16\pi ^2}-
\frac {(-5 +11 \ln (\frac {4}{3}))}{16}
\gamma ^2 (\mu D)^2 \lambda  ^2  \nonumber \\ 
&&\frac {\ln (cm^2a^2)}{(4\pi)^2}
+(-90.5)\gamma ^4 (D\mu)^3\lambda  ^2 \frac {\ln (c m^2a^2)}{(4\pi)^3}.
\label{eq:ggz}
\end{eqnarray}
By using $\delta _0 = \delta +\frac {1}{4}\lambda  ^2 (D \mu)^2 \gamma
^2
\frac {\ln (c m^2a^2)}{(4\pi)^2} $,
we can calculate the recursion relation of $g$, as will be shown
in the next section.
As we will see the term $\ln (c m^2a^2)^2$ is canceled when
we derive the Callan-Symanzik (CS) \cite{Zinn,amit}  equation.
Thus the scaling equation is consistent.

\section { The Calculation of Z$_\lambda $ }
\label{sub5la}
~For the calculation of $Z_{\lambda }$, one should consider the
vertex function $\Gamma _{1,2}$.
For the diagrams in  Fig. 16 and Fig. 17, one can write down the
associated integrals as:
\begin{equation}
\mbox{nla}=\int _{-\infty}^{\infty} d^2 \vec k \int _{-\infty}^{\infty}
d \Omega \frac { 2D\mu ^2 \vec p \cdot (\frac {\vec p}{2}-\vec k)
}{\{\mu[(-\frac {\vec p}{2}-\vec k)^2+m^2
]-i\Omega\}\{\mu ^2(\frac {\vec p}{2}-\vec k)^2+m^2
]^2+\Omega ^2\}} f(\Omega)
\label{eq:laa}
\end{equation}
and
\begin{equation}
\mbox{nlb}=\int _{-\infty}^{\infty} d^2 \vec k \int _{-\infty}^{\infty}
d \Omega \frac {2D\mu ^2\vec p\cdot (\frac {\vec p}{2}-\vec k)
}{\{\mu[(\frac {\vec p}{2}+\vec k)^2+m^2
]-i\Omega\}\{\mu ^2(\frac {\vec p}{2}-\vec k)^2+m^2
]^2+\Omega ^2\}} f(\Omega)  ,
\label{eq:lab}
\end{equation}
where
\begin{eqnarray}
f(\Omega)&=&\gamma  ^2 \int _{-\infty}^{\infty}dt e^{i\Omega t}
[R_0(0,t) e^{\gamma  ^2 C_0(0,t)}]\nonumber  \\
&=& - \int _{-\infty}^{\infty}dt e^{i\Omega t}\frac {1}{\mu ^2 D}
[R_0(0,t)\gamma  ^2] e^{\gamma  ^2 C_0(0,t)} \nonumber  \\
&=& \frac {1}{\mu ^2 D}(i\Omega)\int _{0}^{\infty}
dt e^{i\Omega t}(e^{\gamma  ^2 C_0(0,t)}-1) .
\label{eq:fab}
\end{eqnarray}

For simplicity, let $x=(\frac {\vec p}{2}+\vec k)^2$, and
$y=(\frac {\vec p}{2}-\vec k)^2$.
The summation of   nla in Eq.~\ref{eq:laa} and nlb  in Eq.~\ref{eq:lab} 
is proportional to
$\int _{-\infty}^{\infty} d\Omega \frac {2x}{(x^2+\Omega
^2)(y^2 +\Omega ^2)}f(\Omega)$.

With the help of Eq.~(\ref{eq:fab}),
the frequency dependent part can be integrated out first:
\begin{equation}
\int _{-\infty}^{\infty} d\Omega \frac {(x )(i\Omega)
e^{i\Omega t}}{(x^2+ \Omega ^2)(y^2 +\Omega ^2)}           k
\sim  \frac {e^{-x t}}{x^2-y^2}-\frac {e^{-y t}}{x^2-y^2}
\end{equation} 
 We then  have:
\begin{eqnarray}
\lefteqn{\frac {1}{(2\pi)^2}\int _{-\infty}^{\infty} d^2 k
\frac {\vec p\cdot(\frac {\vec p}{2}-\vec k)}{4(\frac
{p^2}{4}+k^2+m^2)(\vec p\cdot\vec k)\mu ^2}
\{e^{-1\mu [(\frac {\vec p}{2}+\vec k)^2+m^2]t}-
e^{-1\mu [(\frac {\vec p}{2}-\vec k)^2+m^2]t}\}} \nonumber \\
&& \times \mu [(\frac {\vec p}{2}+\vec k)^2 + m^2]\nonumber  \\
&=& \vec p \cdot (\frac {\vec p}{2}-\vec k)
(-2) \frac {e^{-\mu (\frac {p^2}{4}+k^2+m^2)t}t(\vec p\cdot\vec
k)}{(\vec p\cdot\vec k)} +\vec p (\frac {\vec p}{2}-\vec k)
\frac {e^{-\mu (\frac {p^2}{4}+k^2+m^2)t}}{\frac
{p^2}{4}+k^2+m^2}(-2) (\vec p\cdot\vec k)t   \nonumber \\
\label{eq:abf}
\end{eqnarray}
In the hydrodynamic (long wave-length) limit, $\vec p \rightarrow  0$,
The relevant term in the  first term in Eq.~(\ref{eq:abf})
can be easily found as:
\begin{equation}
\int _{-\infty}^{\infty} d^2 \vec k (-2) \frac {p^2}{2}
te^{-\mu (\vec k ^2+m^2)t}=-p^2 \frac {1}{2\mu}e^{-\mu m^2 t} .
\end{equation}
The relevant term in second term of Eq.~(\ref{eq:abf})  is:
\begin{equation}
p^2(2)\int _{0}^{2\pi}d\theta \cos ^2 \theta \int _{0}^{\infty}k
dk t \frac {k^2 e^{-\mu(k^2+m^2)t}}{k^2+m^2}=
p^2 \frac {1}{2\mu} e^{-\mu m^2 t}.
\end{equation}
Thus there are no contributions to the renormalization of $\lambda $
due to their mutual cancellation.
Other possible diagrams arise, but result in no contributions.
In Fig. 18.  those two diagram will not contribute the renormalization
when   one impose the long-time  prescription ($\Omega_{ext} =0$).
The diagrams in Fig. 19.  do not contribute neither  since  the 
interaction $g$
is local in space and therefore there is no  $p$-dependent part  of
the vertex $\Gamma_{1,2}$.

To sum up, $\lambda $ suffers no renormalization within
the perturbative expansion to  order of $g$, and therefore 
$Z_{\lambda }=1$.

\section{ Recursion Relations}
\label{sec77}

Once the Z-factors are known to the leading order in $g$, the
recursion relations are obtained via the so-called 
$\beta $-functions \cite{Zinn,janssen,amit}:

\begin{equation}
\beta_{\mu }=\kappa (\frac{\partial \mu}{\partial \kappa})_b
 =\mu \kappa (\frac {\partial \ln Z_{\tilde{\phi}
}}{\partial \kappa })_{b},
\label{eq:mu}
\end{equation}
\begin{equation}
\beta_{D}=\kappa (\frac{\partial D}{\partial  \kappa})_b 
=-D\kappa (\frac {\partial \ln Z_{D}}{\partial \kappa
})_{b}
,
\label{eq:dd}
\end{equation}

\begin{equation}
\beta _g =\kappa (\frac {\partial g }{\partial \kappa})_b
=-g\kappa (\frac {\partial \ln Z_{g}}{\partial \kappa})_b
\label{eq:gg}
\end{equation}
\begin{equation}  
\beta _{\bar{\nu}}=\kappa (\frac{\partial \bar{\nu}}{\partial \kappa })_b
\end{equation} 
where   subscript $b$   means that all bare parameters are
fixed when one performs the  differentiations \cite{Zinn,janssen,amit} and
 $\kappa$ is a mass scale.
The renormalization of the couplings
may also be related to the same $\beta $ functions.

The renormalization $Z$ factors are the ratios between the 
corresponding renormalized and 
bare parameters. Therefore it is a standard procedure to 
extract from their dependence on the
momentum scale $\kappa$ (or the bare mass $m_0$) the flow 
of the renormalized couplings under rescaling of 
all  length scales by $\tilde{b}=\exp (l)$. The first step 
is to compute the so-called beta functions which
when subtracted from the naive (engineering) dimension of 
the couplings yield the flow equations.
Following this procedure, rescaling length scales 
$x \rightarrow  \tilde{b}  x$, momenta $k \rightarrow  
\tilde{b}^{-1} k$, time
$t \rightarrow  \tilde{b}^z t$ and frequencies 
$ w \rightarrow  \tilde{b}^{-z} w $, we find the 
following recursion relations: 

\begin{eqnarray}
\frac{d\nu}{dl}&=&0  \label{eq:rnu}\\
\frac{d\bar{\nu}}{dl}&=& \frac {\pi\gamma^2}{4\nu (D\mu)^3} g^2 \\
\frac{dF}{dl}&=& 2F+\pi\lambda \label{eq:rf} \\
\frac{dD}{dl}&=&(\frac{\lambda^2}{8\pi}D\mu+
\frac{\gamma^2\sqrt{c}g}{D\mu})D \label{eq:rd} \\
\frac{d\mu}{dl}&=&(-\frac{\gamma^2 \sqrt{c}g}{D\mu})\mu 
\label{eq:rmu} \\
\frac{dg}{dl}&=&(2-\frac{D\mu \gamma^2}{2\pi}-
\frac{\lambda^2 c'}{\gamma^2}
)g-\frac{2\pi}{(D\mu)^2} g^2 \label{eq:rg} \\
\frac{d\lambda}{dl}&=&0 \label{eq:rla}.
\end{eqnarray}

$\gamma$ is not renormalized because $Z_{\phi}=1$ and 
keeps its bare value $\gamma=2\pi$.
For the same reason $\nu$ is not renormalized and 
may be chosen as $\nu =1$.

The two constants are $c=\frac{1}{4} e^{2E}=0.7931$ 
where $E$ is  the Euler Constant.
 and $c' \sim 180.08$, which is derived from the sum 
of the terms contributing to $Z_g$ in 
 Eq.~(\ref{eq:ggz}).

\section { The Asymptotic Behavior}
\label{sectab}

In this chapter we proceed with the analysis of the physical 
implications of the recursion relations. 
We begin, in the next section by looking at the asymptotic 
destination of the flows which will yield 
the 
physical properties on very large scales of time and space. 
In the following subsection will analyze the
crossover behavior which determines the properties on 
intermediate scales.

\underline {1. The asymptotic behavior}

The analysis of the recursion relations may be  
facilitated by the introduction of a "temperature"
like variable (temperature is not well-defined  far 
away from equilibrium where the Einstein 
relation does not hold). Here we define it  by $T=D \mu$ 
(it is not the thermodynamic temperature).
Its recursion relation  is obtained from Eq. ~\ref{eq:rd} 
and Eq. ~\ref{eq:rmu}.It obeys: 
\begin{equation}
\frac{d T}{d l}=\frac{\lambda  ^2}{8 \pi ^2} T,
\end{equation} 
where the critical value $D\mu = 1/\pi$ is substituted.
Since $\lambda $ and $\gamma$ are kept
constant, this equation may be  integrated:
\begin{equation}
T(l)=T_0 e^{\frac {\lambda  ^2}{8 \pi^2}l} .
\label{eq:rtem}
\end{equation}

We see that no matter how small $T_0$ is $T(l)$ will grow 
indefinitely 
with $l=\ln b$ such that:
\begin{equation}
T(l)=T_0 (\frac {L}{a})^{\lambda  ^2 / 8\pi^2}.
\end{equation}

So the effective "temperature" becomes higher on longer 
length scales.
Asymptotically the system is always at a high "temperature". 
The growth of  $T$
is, however, quite slow. Therefore crossover effects discussed 
below plays an important role.

What is the effect of high $T$? For that we have to look at 
the flow of $g$:

\begin{equation}
\frac {dg(l)}{dl}=[2-\frac {T(l)}{2\pi}\gamma^2 -\frac {c^{'}
\lambda }{\gamma ^2}
]g(l) -\frac {\gamma ^4 g^2}{8\pi ^2} .
\end{equation}

It is clear that if $T(l)$ grows very large it will cause $g(l)$ 
to decay to zero,
no matter what are the bare values  $g_0$,  $T_0$ and $\lambda $. 
Once $g\rightarrow  0$ the asymptotic
behavior becomes equivalent to that of the KPZ equation.

We, thus, conclude that asymptotically on very large scales 
and very long time the scaling 
properties are these  associated  with  a far-from equilibrium 
growth without the disorder and
the periodic potential. The behavior will be determined by the  
effect of the  lateral 
growth alone. The KPZ properties in $2+1$  dimensions 
(dominated by an inaccessible
fixed  point) will be the asymptotic ones for the system 
under consideration.

\underline {The Crossover Behavior}
As we have found in the previous section, the "temperature" 
$T(l)$ rises with the scale
quite slowly. Hence the decay of $g$ to zero
might also be slow. As long as $g$ is not vanishing the 
effects of the disorder and the 
periodic potential are still felt. Hence we should expect a 
slowing down of the dynamics.
This slowing down will be observable on larger scales as well 
because the mobility obeys the
equation:
\begin{equation}
\frac{\partial \mu}{\partial l} = -\frac{\gamma ^2 \sqrt{c}
g(l)}{T(l)} \mu (l),
\end{equation}
and therefore

\begin{equation}
\mu (l)= \mu_0 e^{-\gamma ^2 \sqrt{c}\int _{0}^{l}dl^{'}
[g(l^{'})/T(l^{'})] }
\end{equation}.
The ratio $\mu (l) / \mu _0$ does depend on the integral:
\begin{equation}
J(l)=\int _{0}^{l} \frac {g(l ^{'})}{T(l^{'})} dl ^{'}.
\label{eq:rr}
\end{equation}
and clearly $J(l)$ is sensitive to $g(l)$ on small scales as well.

To evaluate $J(l)$ we need to know $T(l)$ given in 
Eq.~\ref{eq:rtem} and $g(l)$ which we calculate next.
Given $T(l)$, $g(l)$ is found by integration of its recursion 
relation:
\begin{equation}
\frac {1}{g(l)}=\frac {1}{g(0)} e^{s(l)}-\frac {\gamma ^4}{8\pi^2}
e^{s(l)} \int _{0}^{l}dx e^{-s(x)},
\end{equation}
where
\begin{equation}
s(x)=[\frac {\lambda  ^2 c^{'}}{\gamma ^2}-2]x +T_0  
\frac {\gamma ^4}{\pi \lambda  ^2}
(e^{\lambda  ^2 x/ 2\gamma ^2}-1)   .
\end{equation}

Clearly the second term dominates for large x. Hence
we see that for large $l$:
\begin{equation}
g(l) \sim \exp \{-\exp (Al)\} 
\end{equation}
with $A\sim \frac{\lambda  ^2}{8 \pi ^2}$.

It is easy to see that for large enough $l$, $g(l)$ decays 
to zero faster than exponential.
Since $T(l)$ diverges, the  most important   contribution to
$J(l)$ comes from small (or at most 
intermediate)  values of $l$. At large $l$,   $J(l)$ approaches 
asymptotically a constant and its dependence on $l$ becomes much weaker.

This asymptotic value of $J(l)$ depends mostly on the bare 
values of the parameters. To summarize,
the mobility decays fast on initial small scales and then 
saturates to an almost constant value on large scales.

\section{Conclusions}

In this work we have investigated the behavior of a 
class of growth systems in which three 
different effects play  important roles: Periodicity, 
disorder and lateral growth. Our model applies 
to the situations in which all three effects are present 
in deposition processes or 
solidification of rigid crystals.

In I we looked at the system near-equilibrium when the 
lateral growth is negligible.
There we found a continuous transition from a rough 
phase at high temperature into a super-rough and glassy
phase for $T<T_g$.

The main conclusion of the present work is that far-from 
equilibrium the KPZ term prevent this transition.
The ultimate asymptotic behavior is dominated by the 
KPZ non-linearity while the second non-linear
term, (obtained upon averaging the periodic potential 
over the disorder) is irrelevant.

We have seen, however, that while this term is decaying 
it still affects the behavior
on intermediate scales. The effect of the disorder 
in the periodic potential is to slow the dynamics.
In particular the mobility decays from its bare value as:

\begin{equation}
\frac{\mu (l)}{\mu (0)} = \exp [-4 \pi ^2\times 1.78 J(l)],
\end{equation} 
where $J(l)$ is given by Eq.~\ref{eq:rr}.

It is clear that the behavior on  intermediate scales 
 drastically depends on the bare values of the 
parameters.

We have also shown that the theory is consistently  
renormalizable to first order in 
$\lambda $ and $g$. To obtain the correct renormalization 
we had to keep diagrams up to
and including three non-trivial loops. As far as we are 
aware of no other calculation 
have shown the renormalizability up to this order for a
 dynamic system for which the 
fluctuation-dissipation theorem is not satisfied. It is 
reassuring  to see that the 
renormalization group can be successfully applied to 
the dynamics far-from equilibrium.

\section*{Acknowledgments}

One of us (Y.-C. T.) wishes  to thank Dr. Z. Yang for  insightful 
discussions.   
Acknowledgment is made to the donors of The Petroleum Research Fund,
administered by the ACS, for support of this research.

\newpage 
\appendix
\section { 2-loop Calculation for $Z_{\lowercase{g}}$}

As explained in Chapter IV, 
the calculation of $Z_g$ is based on the evaluation of the vertex function 
$\Gamma_{2,0}$. 

For clarity, we neglect
the prefactors and symmetry factors.
Here  
we have  6 2-loop and 2 3-loop Feynman diagrams. 
The 2-loop diagrams will be   denoted  by FD 2l$\#$,  
and  3-loop diagrams will  be denoted by FD 3l$\#$, where $\#$ 
  stands  for  the   sequel. 
First we look at the 2-loop diagrams. 
The basic rules \cite{janssen,Domi} for  the calculations of 
these diagrams
have been described in the    Appendix  of our previous paper I.
  Here we shall
simply write down the corresponding integral for each diagram.

We employ the momentum-time representation for correlation
and response functions, in terms of which
the corresponding integrals will be easily handled.

Feynman  diagram (FD) 2l1 is shown in Fig. 8.
The corresponding
integral over time is given by:
\begin{eqnarray}
\lefteqn{\mbox {FD 2l1}} \nonumber \\
&=&\int _{0}^{\infty}\! dt_y \!
\int _{0}^{t_y}\! dt_x \frac{[\vec q\cdot
(\vec q-\vec p)](\vec p\cdot \vec q)}{(q^2+m^2)(p^2+m^2)} 
e^{-[q^2+(\vec p -\vec q)^2+2m^2]t_x}e^{-(p^2+m^2)t_y} 
e^{-(p^2+m^2)(t_y-t_x)}. \nonumber \\
\label{eq:2l1i}
\end{eqnarray}
The integration of the time dependent sectors gives:
\begin{eqnarray}
\lefteqn{\int _{0}^{\infty}dt_y \int _{0}^{t_y}dt_x
e^{-[q^2+(\vec p -\vec q)^2+m^2-p^2]t_x}
e^{-2(p^2+m^2)t_y} } \nonumber  \\
&=& \int _{0}^{\infty} dt_y e^{-2(p^2+m^2)t_y}\frac
{-1}{[q^2+(\vec p -
\vec q)^2+m^2-p^2]}[e^{-[q^2+(\vec p -\vec q)^2+m^2]t_y}-1]
\nonumber  \\
&=& \frac {1}{2(p^2+m^2)[p^2+q^2+(\vec p -\vec q)^2+3m^2]}.
\end{eqnarray}
By  decomposing  $\vec p\cdot\vec q$  into
\begin{equation}
\vec p\cdot\vec q=\frac {-1}{2}[p^2+q^2+(\vec p -\vec q)^2+3m^2-
2p^2-2q^2-3m^2]
\end{equation}
and $\vec q \cdot (\vec q -\vec p) $  into
\begin{equation}
(q^2-\vec p\cdot \vec q)=\frac {1}{2}[p^2+q^2+(\vec p -\vec
q)^2+3m^2-2p^2-3m^2]
\label{eq:2l1dc}
\end{equation}
we obtain
\begin{eqnarray}
\lefteqn{\frac {(\vec q\cdot\vec
p)}{(p^2+m^2)(q^2+m^2)[p^2+q^2+(\vec p -\vec q)^2+3m^2]} }
\nonumber \\
&=&-\frac {1}{2}[\frac {1}{(p^2+m^2)(q^2+m^2)}
-\frac {2}{(p^2+m^2)(q^2+m^2)[p^2+q^2+(\vec p -\vec
q)^2+3m^2]} \nonumber \\
&&-\frac {2}{(p^2+m^2)^2[p^2+q^2+(\vec p -\vec q)^2+3m^2]}
 + \mbox {irrelevant terms} ].
\label{eq:2l1a}
\end{eqnarray}

Since we impose the minimal subtraction (MS) \cite{Zinn,amit} 
scheme, the
unwanted nonsingular parts (finite parts)  will be ignored
in all  calculations in this paper.
In this whole paper, an irrelevant term means a term which
does not contribute to the singular part of the integral.
We   sometimes use equal sign to
represent the equality of the singular parts on both sides.
          
We substitute Eq.~(\ref{eq:2l1dc}) and Eq.~(\ref{eq:2l1a}) 
into Eq.~(\ref{eq:2l1i}), and
obtain:
\begin{eqnarray}
\lefteqn{\mbox {Eq}.~(\ref{eq:2l1i})}  \nonumber \\
&=&-\frac {1}{4} \{ \frac {(q^2-\vec p\cdot \vec
q)}{(p^2+m^2)^2(q^2+m^2)}-\frac {[p^2+q^2+(\vec p -\vec
q)^2+3m^2-2p^2-3m^2]}{(p^2+m^2)(q^2+m^2)[p^2+q^2+(\vec p -\vec
q)^2+3m^2]}\nonumber  \\
&\quad &-\frac {[p^2+q^2+(\vec p -\vec q)^2+3m^2-2p^2-
3m^2]}{(p^2+m^2)^2[p^2+q^2+(\vec p -\vec q)^2+3m^2]}\}
\nonumber  \\
&=&-\frac {1}{4} [\frac {1}{(p^2+m^2)^2}+\frac {\vec p\cdot \vec
q}{(p^2+m^2)(q^2+m^2)}-\frac {1}{(p^2+m^2)(q^2+m^2)}
\nonumber  \\&& + \frac {2}{(q^2+m^2)[p^2+q^2+(\vec p -\vec
q)^2+3m^2]}
-\frac {1}{(p^2+m^2)^2} \nonumber  \\
&&+ \frac {2}{(p^2+m^2)[p^2+q^2+(\vec
p -\vec q)^2+3m^2]}] \nonumber  \\
&=&[ \frac {1}{4}\tilde{B}-\tilde{C}],
\label{eq:2l1ddd}
\end{eqnarray}
where
\begin{eqnarray}
\tilde{B}&=&\frac{1}{(p^2+m^2)(q^2+m^2)}   \\
\tilde{C}&=&\frac {1}{(p^2+m^2)[p^2+q^2+(\vec p -\vec q)^2+3m^2]},
\end{eqnarray}
and the second term in Eq.~(\ref{eq:2l1ddd}) is discarded 
because
it  vanishes after  the integration over the 
momentum variables.

In the same way, the associated integral for Fig. 9 reads:
\begin{eqnarray}
\mbox {FD 2l2}&=& \int _{0}^{\infty} dt_x \int
_{-\infty}^{t_x}dt_y
\frac {[\vec q \cdot(\vec p - \vec q)]}{(q^2+m^2)}
\frac {[\vec q \cdot(\vec p - \vec q)]}{[(\vec p-\vec q)^2+m^2]}
e^{-(p^2+m^2)t_x}\nonumber \\
&&e^{-[(\vec p- \vec q)^2+q^2 +2m^2]|t_y|}
e^{-(p^2+m^2)(t_x-t_y)} .         
\label{eq:2l2i}
\end{eqnarray}
To begin with, we integrate over the time variables,
and that yields:
\begin{eqnarray}
\lefteqn{ \int _{0}^{\infty}dt_x \int _{-\infty}^{0}dt_y
e^{-2(p^2+m^2)t_x}
e^{(p^2+m^2)t_y +[(\vec p -\vec q)^2+q^2+2m^2] t_y}} \nonumber  \\
&&+ \int _{0}^{\infty}dt_x \int _{0}^{t_x}dt_y
e^{-2(p^2+m^2)t_x}
e^{(p^2+m^2)t_y -[(\vec p -\vec q)^2+q^2+2m^2] t_y} \nonumber  \\
&=& \frac {1}{(p^2+m^2)(p^2+q^2+(\vec p -\vec q)^2+3m^2)}.
\label{eq:2l2a}
\end{eqnarray}

By inserting  Eq.~(\ref{eq:2l2a})  into Eq.~(\ref{eq:2l2i}),
we obtain:
\begin{eqnarray}
\lefteqn{\frac {[\vec q \cdot (\vec p - \vec q)][\vec q \cdot
(\vec p - \vec q)]}
{(q^2+m^2)(p^2+m^2)[(\vec p -\vec q)^2+m^2]
[p^2+q^2+(\vec p -\vec q]^2+3m^2)}} \nonumber  \\
&=& \frac {3}{4}\frac {q^2}{(p^2+m^2)[(\vec p -\vec q)^2+
m^2][p^2+q^2+(\vec p -\vec q)^2+3m^2]} \nonumber  \\
&&-\frac {1}{2}\frac {1}{[p^2+q^2+(\vec p -\vec
q)^2+3m^2](q^2+m^2)}\nonumber  \\
&=&[\frac {3}{4}\tilde {A}-\frac {1}{2}\tilde {C}],
\end{eqnarray}
where
\begin{equation}
\tilde {A}= \frac {q^2}{(p^2+m^2)[(\vec p -\vec
q)^2+m^2][p^2+q^2+(\vec p -\vec q)^2+3m^2]}.
\end{equation}

For the Feynman diagram in  Fig. 10, the integral with the time
 variables is written as:
\begin{equation}
\mbox {FD 2l3}= \frac {(\vec q \cdot \vec p)(\vec p\cdot \vec p)}
{(p^2+m^2)(q^2+m^2)}\int _{0}^{\infty}dt_x\int _{0}^{t_x} dt_y
e^{-(p^2+m^2)(t_x-t_y)}e^{[(\vec p -\vec q)^2  +q^2+2m^2]t_y}.
\label{eq:2l3i}
\end{equation}
Again, the time dependent sectors are integrated out in advance.
\begin{eqnarray}
\lefteqn{\int _{0}^{\infty}dt_x \int _{0}^{t_x}dt_y
e^{-2(p^2+m^2)t_x} e^{-[q^2+(\vec p -\vec
q)^2+2m^2]t_y+(p^2+m^2)t_y}}\nonumber  \\
&=& \frac {1}{2}\frac {1}{(p^2+m^2)[p^2+q^2+(\vec p -\vec
q)^2+3m^2]}.
\label{eq:2l3a}
\end{eqnarray}
With the help of Eq.~(\ref{eq:2l3a}), Eq.~(\ref{eq:2l3i}) is simplified
into:
\begin{eqnarray}
&&\frac {(\vec p\cdot\vec q)(\vec p\cdot\vec
q)}{(p^2+m^2)(q^2+m^2)}
\frac {1}{2}\frac {1}{(p^2+m^2)[p^2+q^2+(\vec p -\vec q)^2+3m^2]}
\nonumber  \\
&& = \frac {1}{4}[4\tilde {C}-\tilde {B}].
\end{eqnarray}

Now we turn to  the diagram in Fig. 11. The associated integral is
represented  by:   
\begin{eqnarray}
\mbox {FD 2l4} &=&-\frac {p^2[(\vec p-\vec q)\cdot \vec
q]}{(q^2+m^2)[(\vec p - \vec q)^2+m^2]}
\int _{0}^{\infty}dt_x \int _{-\infty}^{t_x} dt_y
e^{-(p^2+m^2)t_x} \nonumber \\
&& e^{-[q^2+(\vec p -\vec q)^2+2m^2]|t_y|}
e^{-(p^2+m^2)(t_x-t_y)} .
\end{eqnarray}
The time dependent sectors in Fig.  11 is identical to Eq.~(\ref{eq:2l2a}),
so we will not repeat the  calculation here.

In the same fashion, the integrand takes the form:
\begin{eqnarray}
&&\frac {[q^2-\vec p\cdot\vec q]}{(q^2+m^2)[(\vec p -\vec
q)^2+m^2][p^2+q^2+(\vec p -\vec q)^2+3m^2]} \nonumber  \\
&&= [\frac {1}{2} \tilde {B}-\tilde {A}] .
\end{eqnarray}

For the diagram in Fig. 12, we have the associated integral as:  
\begin{eqnarray}
\lefteqn{\mbox {FD 2l5}}   \nonumber \\
&=&\frac{-[\vec q\cdot (\vec p-\vec
q)](p^2)}{(q^2+m^2)(p^2+m^2)}\!
\int _{0}^{\infty}\! dt_x\! \int _{0}^{t_x}\! dt_y e^{-[q^2+(\vec p -\vec
q)^2+2m^2]t_y}e^{-(p^2+m^2)(t_x-t_y)}e^{-(p^2+m^2)t_x}. \nonumber \\
\end{eqnarray}
The time dependent term is integrated out first:
\begin{eqnarray}
\lefteqn{\int _{0}^{\infty}dt_x \int _{0}^{t_x}dt_y e^{-[q^2+(\vec
p -\vec q)^2+2m^2]t_y}e^{-(p^2+m^2)(t_x-t_y)}e^{-(p^2+m^2)t_x}}
\nonumber  \\
&=& \frac {1}{2} \quad  \frac {1}{(p^2+m^2)[p^2+q^2+(\vec p -\vec
q)^2+3m^2]}.
\end{eqnarray}                       
Then we  obtain:
\begin{equation}
-\frac{[\vec q \cdot(\vec p-\vec q)](p^2)}{(q^2+m^2)(p^2+m^2)}
\frac{1}{2} \frac{1}{(p^2+m^2)[p^2+q^2+(\vec p -\vec q)^2+3m^2]}
=\frac {1}{4}(\tilde {B}-2\tilde {C}) .
\end{equation}

 We turn to the simplest figure  among 2-loop diagrams,
Fig. 13, of which the related integral is
evaluated as:  
\begin{equation}
\{   \int _{-\infty}^{\infty}
\int _{0}^{\infty} dt p^2  R_0(\vec p,t )C_0 (\vec p, t) \}^2
=\frac{1}{4} [\int _{-\infty}^{\infty}
d^d \vec p \frac {1}{(p^2+m^2)}]^2  .
\end{equation}

\begin{eqnarray}
\mbox {FD 2l6}&=& -\frac {1}{4}\int _{-\infty}^{\infty}
d^d \vec p \frac {1}{(p^2+m^2)}   \int _{-\infty}^{\infty}
d^d \vec p \frac {1}{(p^2+m^2)}  \nonumber  \\
&=& -\frac {1}{4}B
\end{eqnarray}

Now we go on to perform the integration over the momentum
variables.
Before the evaluation of  the singular parts of integrals,
it will be helpful to present some identities, which will
play   important roles in later calculations
and have been frequently employed in these types of
calculations. The first one is the Feynman parameterization
formula, which reads:
\begin{eqnarray}
\frac {1}{A^{\alpha}B^{\beta}\cdots E^{\sigma}}&=&
\frac {\Gamma (\alpha +\beta +\gamma  +\cdots+\epsilon )}{\Gamma (\alpha)
\Gamma (\beta)\Gamma (\gamma )\cdots\Gamma (\sigma)}\int_{0}^{1}
\cdots\int _{0}^{1}dx dy dz \cdots
\delta(1-x-y-z\cdots)  \times
\nonumber  \\
&&\frac {x^{\alpha -1}y^{\beta -
1}\cdots z^{\sigma -1}}
{(Ax+By+\cdots +Ez)^{\alpha + \beta +\gamma  +\cdots +\sigma }} .
\end{eqnarray}
A set of integral formulas  is also valuable and is  shown as below:
\begin{equation}
J_0=\int _{-\infty}^{\infty} d^d \vec k \frac {1}{(\vec k ^2+
2\vec k \cdot \vec p +M)^{\alpha}}=
\frac {\pi ^{d/2}}{\Gamma (\alpha)}(M-p^2)^{d/2-\alpha}
\Gamma (\alpha - \frac {d}{2}).
\end{equation}
\begin{equation}
\int _{-\infty}^{\infty} d^d \vec k  \frac {k^{\nu}}{(\vec k ^2+
2\vec k \cdot \vec p +M)^{\alpha}}=-p^{\nu}J_0.
\end{equation}
With these formulas, one can evaluate the singular
parts of  the integrals.
Let $X=\int\int d^d \vec p d^d \vec q \tilde {X}$, where
$X=A,~ B$ or $C$, and $d=2+\epsilon  $.
The evaluations of  $B,C$ and $A$ are carried out as below:
\begin{eqnarray}
B^{1/2}&=&\int _{-\infty}^{\infty} d^d \vec p \frac
{1}{(p^2+m^2)}     \nonumber \\
&=&\pi ^{d/2}(m^2)^{\frac {d}{2}-1} \frac {\Gamma (1-\frac
{d}{2})}{\Gamma (1)}  \\
C&=& \int _{-\infty}^{\infty} d^d \vec p \int _{-\infty}^{\infty}
d^d \vec q  \frac {1}{(p^2+m^2)[p^2+q^2+(\vec p -\vec q)^2+3m^2]}
\nonumber  \\
&=& \frac {1}{2}\int _{-\infty}^{\infty} d^d \vec p
\int _{-\infty}^{\infty} d^d \vec q \frac {1}{(p^2+m^2)}
\frac {1}{(p^2+q^2-\vec p\cdot\vec q+\frac
{3}{4}m^2)}+\mbox {non-singular terms }\nonumber  \\
&=&\frac {1}{2} \int _{-\infty}^{\infty} d^d \vec p
\frac {1}{(p^2+m^2)} \pi ^{d/2} \frac {\Gamma (1-d/2)}{\Gamma
(1)}
\frac {1}{(\frac {3}{4}p^2+\frac {3}{4}m^2)^{-\epsilon   /2}}\nonumber
\\ &=& \frac {1}{2}\pi ^{d} \frac {\Gamma (-\epsilon   /2)\Gamma (-
\epsilon  )}{\Gamma (1-\epsilon   /2)}(\frac {3}{4})^{\epsilon    
/2} m^{2\epsilon   }  \\
A&=& \int _{-\infty}^{\infty} d^d \vec p \int _{-\infty}^{\infty}
d^d \vec q \frac {q^2}{(p^2+m^2)[(\vec p -\vec
q)^2+m^2][p^2+q^2+(\vec p -\vec q)^2+3m^2]} \nonumber  \\
&=& \int _{-\infty}^{\infty} d^d \vec p \int _{-\infty}^{\infty}
d^d \vec q \frac {q^2}{(p^2+m^2)[(\vec p -\vec
q)^2+m^2][p^2+q^2+(\vec p -\vec q)^2+2m^2]}
\nonumber  \\
&=& \frac {1}{2} \int _{-\infty}^{\infty} d^d \vec p
\int _{-\infty}^{\infty} d^d \vec q {\int _{0}^{1} \int _{0}^{1}}
_{x+y<1}dx dy \nonumber  \\
&&\frac {\Gamma(3)q^2}{\{(1-x-y)(p^2+m^2)+x[(\vec p -\vec
q)^2+m^2]+y(p^2+q^2-\vec p\cdot\vec q +m^2)\}^3} 
\nonumber  \\
&=&\int _{-\infty}^{\infty} d^d \vec p
\int _{-\infty}^{\infty} d^d \vec q {\int _{0}^{1} \int _{0}^{1}}
_{x+y<1}dx dy \frac {q^2}{[p^2-\vec p\cdot\vec q(2x+y)
+(x+y)q^2+m^2]^3} 
\nonumber  \\
&=& \int _{-\infty}^{\infty} d^d \vec q q^2   {\int _{0}^{1} \int
_{0}^{1}}_{x+y<1}dx dy \frac {\Gamma (3-d/2)}{\Gamma
(3)}(\pi)^{d/2}\frac {1}{\{ [(x+y)-(x+\frac {y}{2})^2]q^2+m^2
\}^{2-\epsilon   /2}} \nonumber  \\
&=& (\pi)^{d/2}\frac {\Gamma (3-d/2)}{\Gamma (3)}{\int
_{0}^{1}\int _{0}^{1}}_{x+y<1} dx dy  \int _{-\infty}^{\infty}
d^d \vec q  \frac {q^2}{[(x+y)-(x+\frac {y}{2})^2]^{2-\epsilon   /2}
}  \times \nonumber  \\
&& \frac {1}{\{q^2 + \frac {m^2}{[(x+y)-(x+\frac
{y}{2})^2]}\}^{2-\epsilon   /2}}
\nonumber  \\
&=& (\pi)^{d/2}\frac {\Gamma (3-d/2)}{\Gamma (3)}{\int
_{0}^{1}\int _{0}^{1}}_{x+y<1} dx dy  \frac {1}{[(x+y)-(x+\frac
{y}{2})^2]^{2-\epsilon   /2}} \times \nonumber  \\
&& \{m \frac {1}{[(x+y)-(x+\frac
{y}{2})^2]^{1/2}}\}^{d+2-2(2-\epsilon   /2)} (\pi)^{d/2}
\frac {\Gamma (1+d/2)}{\Gamma (d/2)}
\frac {\Gamma (-\epsilon   )}{\Gamma (2-\epsilon   /2)}\nonumber  \\
&=& (\pi)^{d}\frac {\Gamma (2+\epsilon   /2)}{\Gamma (1+\epsilon   /2)}
\frac {\Gamma (-\epsilon   )}{2} {\int _{0}^{1}\int _{0}^{1}}_{x+y<1} dx
dy m^{2\epsilon   } \frac {1}{[(x+y)-(x+\frac {y}{2})^2]^{2+\epsilon   /2}}.
\label{eq:2la}
\end{eqnarray}

The singular parts of integral in Eq.~(\ref{eq:2la}) can be evaluated
by the changes of variables, $x=st$, $y=s(1-t)$.
Let $ A ^{'}$ denote the $x,y$ dependent part in Eq.~(\ref{eq:2la}).
We obtain
\begin{eqnarray}
A'&=& {\int _{0}^{1}\int _{0}^{1}}_{x+y<1} dx dy
\frac {1}{[(x+y)-( x+\frac {y}{2})^2]^{2+\epsilon   /2}}\nonumber  \\
&=&\int _{0}^{1}\int _{0}^{1} dsdt \frac {s}{[s-s^2(1-\frac
{t}{2})]^{2+\epsilon   /2}}\nonumber  \\
&=& \int _{0}^{1}\int _{0}^{1}ds dt \frac {1}{s^{1+\epsilon   /2}[1-(1-
\frac {t}{2})s]^{2+\epsilon   /2}} \nonumber  \\
&=& \int _{0}^{1}\int _{0}^{1}ds dt \frac {1}{s^{1+\epsilon   /2}}
\{\frac {1}{[1-(1-\frac {t}{2}s)]^{2+\epsilon   /2}}-1\}
+\int _{0}^{1}ds \frac {1}{s^{1+\epsilon   /2}}.
\label{eq:2lap}
\end{eqnarray}
>From the form of Eq.~(\ref{eq:2lap}),
it is not hard to see
that the singular term  as $ \epsilon   \rightarrow  0$ is  the manifestation
of the singular behavior of the pole $s=0$ in the integral.
To simplify the expressions of equations, the term $1-\frac {t}{2}$ is
symbolized by $\alpha$.
Eq.~(\ref{eq:2lap}) can be rewritten as:
\begin{equation}
\mbox{Eq}.~(\ref{eq:2lap}) = \int _{0}^{1}\int _{0}^{1}ds dt 
\frac {1}{s^{1+\epsilon 
 /2}}
\frac {2\alpha s - \alpha ^2 s^2}{(1-\alpha s)^{2+\epsilon   /2}}
+\int _{0}^{1}ds \frac {1}{s^{1+\epsilon   /2}}.
\label{eq:2lapo}
\end{equation}

The analysis of two terms in Eq.~(\ref{eq:2lapo}) will 
be carried out in order.
The first term in Eq.~(\ref{eq:2lapo})  is recast into:
\begin{eqnarray}
&&\int _{0}^{1}\int _{0}^{1}ds dt
\frac {1}{s^{\epsilon  }}\frac {\alpha}{(1-\alpha s)^{2+\epsilon   /2}}
+\int _{0}^{1}\int _{0}^{1}ds dt
\frac {1}{s^{\epsilon  }}\frac {\alpha}{(1-\alpha s)^{1+\epsilon  
 /2}}\nonumber
\\ &=& \int _{0}^{1}\int _{0}^{1}ds dt \frac {\alpha}{(1-\alpha
s)^{2+\epsilon   /2}}+ \epsilon   \int _{0}^{1}\int _{0}^{1}
dsdt \frac {[\ln (s)] \alpha}
{(1-\alpha s)^{2+\epsilon   /2}} \nonumber \\ &&
+\int _{0}^{1}\int _{0}^{1}ds dt \frac {\alpha}{(1-\alpha
s)^{1+\epsilon   /2}}+O(\epsilon   ).
\label{eq:2lapt}
\end{eqnarray}
The second term in Eq.~(\ref{eq:2lapt}) is of order $\epsilon  
$ and therefore is discarded.
The first term in Eq.~(\ref{eq:2lapt}) is evaluated as:
\begin{eqnarray}
&&\int _{0}^{1}\int _{0}^{1}ds dt \frac {\alpha}{(1-\alpha
s)^{2+\epsilon   /2}} = \frac {1}{1+\epsilon   /2}\{\int _{0}^{1}
dt[\frac {1}{[1-
\alpha ]^{1\epsilon   + /2}}-1]\} \nonumber  \\
&&=\frac {1}{1+\epsilon   /2}\{\int _{0}^{1}dt \frac {1}{(t-\frac
{t^2}{4})^{1+\epsilon   /2}}-1\},
\end{eqnarray}
where
\begin{eqnarray}
\int _{0}^{1} dt \frac {1}{(t-\frac {t^2}{4})^{1+\epsilon   /2}}
&=&\int _{0}^{1}dt \frac {1}{t^{1+\epsilon   /2}}[\frac {1}{(1-\frac
{t}{4})^{1+\epsilon   /2}}-1] +\int _{0}^{1}dt \frac {1}{t^{1
+\epsilon   /2}}
\nonumber  \\
&=& \int _{0}^{1}dt \frac {1}{t^{1+\epsilon   /2}} \frac {\frac
{t}{4}+(1-\frac {t}{4})\frac {\epsilon   }{2} \ln (1-\frac {t}{4})+
\cdots }{(1-\frac {t}{4})^{1 +\epsilon   /2}}
+\int _{0}^{1}dt \frac {1}{t^{1+\epsilon   /2}}\nonumber  \\
&=& \int _{0}^{1}dt \frac {1}{4 t^{1+\epsilon   /2}}\frac {1}{(1-\frac
{t}{4})^{1+\epsilon   /2}}+O(\epsilon   )+\frac {1}{-\epsilon  /2}
\nonumber  \\
&=&\frac {1}{4}\int _{0}^{1}dt \frac {1}{(1-\frac {t}{4})}
-\frac {2}{\epsilon   } +O(\epsilon   )\nonumber  \\
&=&\ln (\frac {4}{3}) -\frac {2}{\epsilon   }+O(\epsilon   ) .
\end{eqnarray}

The third term in Eq.~(\ref{eq:2lapt}) is:
\begin{eqnarray}
\int _{0}^{1}\int _{0}^{1}ds dt \frac {\alpha}{(1-\alpha s)^{1
+\epsilon   /2}}&=&\int _{0}^{1}\int _{0}^{1}ds dt \frac {\alpha}
{(1-\alpha
s)}+O(\epsilon   ) \nonumber  \\
&=&-\int _{0}^{1}dt\ln [1-(1-\frac {t}{2})^{2}] +O(\epsilon   ) \nonumber \\
&=&3\ln (\frac {4}{3})+O(\epsilon   )
\end{eqnarray}

The second term in Eq.~(\ref{eq:2lapo}) equals to $-\frac {2}{\epsilon  }$.

With the combination of the prefactors in Eq.~(\ref{eq:2la}) and the
results obtained above, the singular part of $A$ is given by:
\begin{equation}
A=(\pi)^{d}[\frac {2}{\epsilon   ^2}-\frac {2}{\epsilon   }\ln (\frac {4}{3})
-\frac {2\gamma  }{\epsilon   }+\frac {1}{\epsilon   }]+ 
\mbox { finite terms } ,
\end{equation}
where $\gamma $ is the Euler number.
The finite part of the subdivergence diagram can be neglected, if the
scale equation is well defined. One can always scale it away.\cite {Zinn}
In the  short distance cutoff \cite{Zinn,GS}, they appears as:
\begin{eqnarray}
A&=&\frac {1}{16\pi ^2}[\frac {1}{2}(\ln (cm^2a^2))^2-
\ln (\frac {4}{3})\ln (cm^2a^2)+\frac {1}{2}\ln (cm^2a^2)] \\
B&=&\frac {1}{16\pi ^2}[(\ln (cm^2a^2))^2]    \\
C&=&\frac {1}{16 \pi ^2}[\frac {1}{4}(\ln (cm^2a^2))
+ \frac {1}{4}\ln (\frac {3}{4})\ln (cm^2a^2)]
\end{eqnarray}

Now we retrieve the symmetry factors for each 2-loop diagrams (see
Table~\ref{t2x}).
\begin{table}
\begin{center}
\caption{{\bf Symmetry factors and integrals of 2-loop diagrams}}
\vspace{0.5in}
\begin{tabular}{|c|c|c|} \hline
&& \\
Feynman diagrams  &Symmetry factors & Integrals  \\
&& \\ \hline &&\\
FD.2l1  &16 &($\frac {1}{4}$B-C) \\  &&\\
FD.2l2  &8  &($\frac {3}{4}$A-$\frac {1}{2}$C) \\ &&\\
FD.2l3  &32 &(C-$\frac {1}{4}$B)    \\ &&\\
FD.2l4  &16 &($\frac {1}{2}$B-A)       \\   &&\\
FD.2l5  &32 &($\frac {1}{4}$B-$\frac {1}{2}$C) \\  &&\\
FD.2l6  &16 & -$\frac {1}{4}$B \\   &&\\
\hline
\end {tabular}
\end{center}
\label{t2x}                                    
\end{table}
As one can see from Fig. 8 -- Fig. 13,
only Fig. 9, Fig. 12, and
Fig. 13  contain  sub-divergent diagrams.
One also can verify this   from the results listed in  Table~\ref{t2x}.
The leading terms in
the diagrams  without sub-divergences,
such as  Fig. 8,
Fig. 10, and Fig. 11, are  of order $\ln(cm^2a^2)$.
On the other hand, the leading terms of the diagrams with
subdivergence as mentioned above are   of order $[\ln(cm^2a^2)]^2$.
Furthermore, the leading divergences of the diagrams in  Fig. 12
and Fig. 13 are
canceled out by each other. They have the same type of sub-divergences,
(see the subdiagrams enclosed   by the boxes in their own figures)
which are not present in the lower order (1-loop) calculation.
Another diagram in Fig. 9 includes the sub-divergent diagram (see the 
subdiagram  enclosed  by the box in Fig. 9)
which occurs in the 1-loop calculation for $D$. As usual, it will be
canceled when  one calculates the  recursion relations, eventhough the $Z_g$
factors  contain some terms  like $[\ln (cm^2)a^2 ]^2$.
As we will show in the next Appendix, the 3-loop diagrams
do not consist of any sub-divergent diagrams. Thus,
the subdivergence in the present expansion only occurs in 2-loop diagrams.
The cancellation of sub-divergences in Fig.  12 and Fig. 13, 
and    
that of  Fig. 9 and Fig. 3 
ensure the renormalizability of
this theory (at least at this order). 
The sum of leading divergences and sub-leading divergences are
summed up to contribute to $Z_g$  (see 
the third and fourth terms in Eq.~\ref{eq:ggz}).

\section { 3-loop Calculation  for $Z_{\lowercase{g}}$ }

In this Appendix, we shall present  the calculations for 3-loop diagrams,
which are mentioned in Chapter IV.

The diagram  shown in Fig. 14 representing  the  integral is given by:
\begin{eqnarray}
\lefteqn{\mbox {FD 3l1}}\nonumber \\
&=& \int _{-\infty}^{\infty} d^d \vec p
\int _{-\infty}^{\infty} d^d \vec q
\int _{-\infty}^{\infty} d^d \vec k
\frac {-[\vec p\cdot\vec q][(\vec p-\vec k)\cdot\vec k]}
{(q^2+m^2)[(\vec p-\vec k)^2+m^2](k^2+m^2)}\times \nonumber \\
&&\int _{0}^{\infty}dt_y
\int _{0}^{t_y}dt_x e^{-[(q^2+m^2)+[(\vec p-\vec q)^2+m^2]](t_y-
t_x)}e^{-(p^2+m^2)t_x}
e^{-[(k^2+m^2)+[(\vec p-\vec k)^2+m^2]]t_y}. \nonumber \\
\label{eq:3l1i}
\end{eqnarray}

To simplify the calculation, we denote
$a=(q^2+m^2)$, $b=[(\vec p-\vec q)^2+m^2]$, $c=(k^2+m^2)$,
$d=[(\vec p-\vec k)^2+m^2]$, and $e=(p^2+m^2)$.
Along the  same line as our  preceding  calculations of  2-loop
diagrams,
we integrate out the time dependent term first.
\begin{eqnarray}
&&\int _{0}^{\infty}\!dt_y
e^{-\{(q^2+m^2)+[(\vec
p-\vec q)^2+m^2]+[(k^2+m^2)+[(\vec p-\vec k)^2+m^2]\}t_y}
\int_{0}^{t_y}\!dt_x\nonumber \\
&&\quad e^{-\{(p^2+m^2)-(q^2+m^2)-[(\vec p-\vec q)^2+m^2]\}t_x}
\nonumber  \\
&&=\int _{0}^{\infty}dt_y e^{-(a+b+c+d)t_y}[e^{-(e-a-b)t_y}-1]
\frac {1}{b+c-e}\nonumber  \\
&&=\int _{0}^{\infty}dt_y \frac {1}{b+c-a}\{e^{-(e+c+d)t_y}-
e^{-(a+b+c+d)t_y}\}\nonumber  \\
&&= \frac {1}{(c+d+e)(a+b+c+d)} .
\end{eqnarray}

Therefore
\begin{eqnarray}
\lefteqn{\mbox{Eq}.~(\ref{eq:3l1i})} \nonumber \\
&=& \int _{-\infty}^{\infty} d^d \vec p
\int _{-\infty}^{\infty} d^d \vec q \int _{-\infty}^{\infty} d^d
\vec k \frac {-[\vec p\cdot\vec q][\vec p -\vec k]\cdot\vec k}
{(q^2+m^2)[(\vec p-\vec k)^2+m^2](k^2+m^2)} \nonumber  \\ &&
\frac {1}{(c+d+e)(a+b+c+d)}\nonumber  \\
&=&(\frac {-1}{2})\int _{-\infty}^{\infty} d^d \vec p
\int _{-\infty}^{\infty} d^d \vec q \int _{-\infty}^{\infty} d^d
\vec k \frac {\vec p\cdot\vec q}{(q^2+m^2)(k^2+m^2)[(\vec p-\vec
k)^2+m^2]} \nonumber \\
&&\frac {1}{\{(q^2+m^2)+(k^2+m^2)+[(\vec p-\vec q)^2+m^2]+[(\vec
p-\vec k)^2+m^2]\}}\nonumber  \\
&&- \frac {2 \vec p\cdot\vec q}{(q^2+m^2)[(\vec p-\vec
k)^2+m^2]\{(k^2+m^2)+(p^2+m^2)+[(\vec p-\vec
k)^2+m^2]\}} \times \nonumber \\
&&\frac {1}{\{(q^2+m^2)+(k^2+m^2)+[(\vec p-\vec q)^2+m^2]+[(\vec
p-\vec k)^2+m^2]\}} \nonumber  \\
&&-  \frac {2 \vec p\cdot\vec q}{(q^2+m^2)(k^2+m^2)[(p^2+m^2)+
(k^2+m^2)+(\vec p-\vec k)^2+m^2]} \times  \nonumber \\
&&\frac {1}{\{(q^2+m^2)+(k^2+m^2)+[(\vec p-
\vec q)^2+m^2]+[(\vec p-\vec k)^2+m^2]\}} ,
\label{eq:3l1a}
\end{eqnarray}
where we have used the identities below to simplify
the expression of the  equation:
\begin{equation}
\frac {\vec s\cdot\vec k}{c+d+e}=\frac {1}{2}[1-\frac {-2k^2-
2s^2-3m^2}{c+d+e}]\frac {1}{a+b+c+d} ,
\end{equation}
where $\vec s=\vec p-\vec k$.
To simplify  the calculation, we treat  Eq.~(\ref{eq:3l1a})
as a linear combination of 3 integrals, 3.1-1, 3.1-2, and 3.1-3.
Namely,
\begin{equation} 
\mbox{Eq}.~(\ref{eq:3l1a})=\frac {-1}{2}[(3-1.1)-2(3-1.2)-2(3-1.3)].
\end{equation} 
The  term  denoted by $3.1-1$  yields:
\begin{eqnarray}
\lefteqn{3.1-1}\nonumber \\
&=&\int _{-\infty}^{\infty} d^d \vec p
\int _{-\infty}^{\infty} d^d \vec q \int _{-\infty}^{\infty} d^d
\vec k \frac {\vec p \cdot \vec q}{(q^2+m^2)(k^2+m^2)[(\vec
p-\vec
k)^2+m^2]}\nonumber \\
&&\frac {1}{\{(q^2+m^2)+(k^2+m^2)+[(\vec p-\vec q)^2+m^2]+[(\vec
p-\vec k)^2+m^2]\}}\nonumber  \\
&=& \int _{-\infty}^{\infty} d^d \vec p\int _{-\infty}^{\infty}
d^d \vec q \frac {1}{2}\quad \frac {\Gamma (1+1+1)}{\Gamma
(1)\Gamma
(1)\Gamma (1)} \quad {\int _{0}^{1} \int _{0}^{1}}_{x+y<1}dx dy
\int _{-\infty}^{\infty} d^d \vec k \nonumber \\
&&\frac {\vec p\cdot\vec q}
{[k^2+\vec k\cdot\vec p (-2x-y)+p^2(x+y)+\vec p\cdot\vec q (-y)+
y q^2+m^2]^{3} }\nonumber  \\
&=& (\pi)^{d/2}\frac {\Gamma (2-\epsilon   /2)}{\Gamma (3)}\frac {\Gamma
(3)}{2}\int _{-\infty}^{\infty} d^d \vec p\int _{-
\infty}^{\infty} d^d \vec q {\int _{0}^{1}\int
_{0}^{1}}_{x+y<1}dx dy  \nonumber  \\
 && \frac {\vec p\cdot\vec q}
{\{[(x+y)-(x+\frac {y}{2})^{2}]p^2+\vec p\cdot\vec q (-y)
+yq^2+m^2\}^{2-\epsilon   /2}}\nonumber  \\
&=& (\pi)^{d/2}\frac {\Gamma (2-\epsilon   /2)}{2} \int
_{-\infty}^{\infty}
d^d \vec p\int _{-\infty}^{\infty} d^d \vec q {\int _{0}^{1}\int
_{0}^{1}}_{x+y<1}dx dy \nonumber \\
&&\frac {\vec p\cdot\vec q }{(q^2+m^2)y^{2-
\epsilon   /2}}\frac {1}{[q^2-\vec p\cdot\vec q +\frac {\Delta}{y}p^2
+\frac {1}{y}m^2]^{2-\epsilon   /2}}\nonumber  \\
&=& (\pi)^{d/2}\frac {\Gamma (2-\epsilon   /2)}{2} \int
_{-\infty}^{\infty}
d^d \vec p\int _{-\infty}^{\infty} d^d \vec q {\int _{0}^{1}\int
_{0}^{1}\int _{0}^{1}}_{x+y<1}dx dy dz \nonumber \\
&&\frac {\Gamma (3-\epsilon  
/2)}{\Gamma (1)\Gamma (2-\epsilon   /2)}
\frac {z^{1-\epsilon   /2} \quad \vec p\cdot\vec q }{y^{2-\epsilon 
/2}[q^2-z\vec
p\cdot\vec q +\frac {z \Delta
}{y}p^2+\frac {z}{y}m^2]^{3-\epsilon   /2}}
\nonumber  \\
&=& (\pi)^{d/2}\frac {\Gamma (2-\epsilon 
 /2)\Gamma (3-\epsilon 
 /2)}{\Gamma
(2-\epsilon 
 /2)}\frac {z^{1 -\epsilon   /2}}{y^{2-\epsilon 
 /2}}\frac {\frac {z}{2}\cdot p^2}{\{[\frac
{z\Delta}{y}-(\frac
{z}{2})^{2}]p^2+\frac {z}{y}m^2\}^{2- \epsilon 
}} (\pi)^{d/2}
\frac {\Gamma (2- \epsilon 
)}{\Gamma (3- \epsilon 
/2)}\nonumber  \\
&=&\frac {(\pi)^d\Gamma (2-\epsilon 
)}{2\times 2}
{\int _{0}^{1}\int _{0}^{1}\int _{0}^{1}}_{x+y<1}dx dy dz
 \int _{-\infty}^{\infty} d^d \vec p \frac {z^{2-\epsilon   /2}}{y^{2-
\epsilon  
/2}(\Box)^{2-\epsilon   }} \frac {p^2}{[p^2+\frac {zm^2}{y \Box 
} ]^{2-\epsilon 
}}\nonumber  \\
&=& \frac {\pi ^d \Gamma (2-\epsilon  )}{4} {\int _{0}^{1}\int
 _{0}^{1}\int
 _{0}^{1} }_{x+y<1} dx dy dz \frac {z^{2-\epsilon   /2}}{y^{2-\epsilon 
 /2}\Box
 ^{2-\epsilon  }}[m(\frac {z}{y\Box })^{1/2}]^{3\epsilon   }
 (\pi)^{d/2} \nonumber \\
&&\frac {\Gamma (2+\epsilon   /2)}{\Gamma (1+\epsilon   /2)}
 \frac {\Gamma (-3/2 \epsilon   )}{\Gamma (2-\epsilon   )}\nonumber  \\
&=&\frac {\pi ^{3/2 d}}{4}\frac {\Gamma (2+\epsilon   /2)\Gamma (-3/2
\epsilon   
 )}{\Gamma (1+ \epsilon   /2)}(m^2)^{3\epsilon   /2} 
{\int _{0}^{1}\int _{0}^{1}\int
_{0}^{1}}_{x+y<1}dx dy dz \frac {z^{2+\epsilon 
 }}{y^{2+ \epsilon   } (\Box)
^{2+\epsilon   /2}} , \nonumber \\
\label{eq:3l11a}
\end{eqnarray}
where   $\Delta = (x+y)-(x - \frac {y}{2}) ^2$  and
$\Box = \frac {z}{y}\Delta -\frac{z ^2}{4}$.

The working principles of extracting    the singularity of the integral 
is based on the  separation of the singular contributions  from 
different points.  All the calculation here  follow  this scenario. 
However, one should be able to keep track of  
those  highly nested procedures. 
In the last line of Eq.~(\ref{eq:3l11a}),
the prefactors before the integral contains a
leading singular-pole of order $\frac {1}{\epsilon   }$,
and thus we should extract the
contribution up to the zero order in  $\epsilon   $ from this integral.
The extraction of the poles and finite parts of this integral
 proceeds as below:
\begin{eqnarray}
\lefteqn{\int _{0}^{1} {\int _{0}^{1}\int _{0}^{1}}_{x+y<1}dx dy dz
\frac {z^{2+\epsilon   }}{y^{2+\epsilon   }\Box ^{2+\epsilon   /2}} }
\nonumber  \\
&=& \int _{0}^{1}dz {\int _{0}^{1}\int _{0}^{1}}_{x+y<1}dx dy
\frac {z^{\epsilon   /2}}{y^{\epsilon   /2}\{[(x+y)-(x+\frac {y}{2})^{2}]-
\frac {zy}{4}\}^{2+\epsilon   /2}}
\label{eq:3l11b}
\end{eqnarray}
Rewriting $x,y$ as $y=st, x=s(1-t)$,
we inherit the simplified equation:
\begin{eqnarray}
\mbox{Eq}.~(\ref{eq:3l11b})&=& \int _{0}^{1}dz \int _{0}^{1}
\int _{0}^{1}ds dt
\frac {s z^{\epsilon   /2}}{s^{\epsilon   /2}t^{\epsilon 
  /2 }[s-s^2 (1-\frac {t}{2})^2
-\frac {z}{4}st]^{2+\epsilon   /2}}\nonumber  \\
&=& \int _{0}^{1}dz \int _{0}^{1}\int _{0}^{1}ds dt
\frac {z^{\epsilon   /2}}{t^{\epsilon   /2}}\frac {1}{s^{1+\epsilon 
}[1-s(1-\frac
{t}{2})^2-\frac {zt}{4}]^{2+\epsilon   /2}} .
\label{eq:3l11c}
\end{eqnarray}
Recast Eq.~(\ref{eq:3l11c}) into:
\begin{equation}
\mbox{Eq}.~(\ref{eq:3l11c})=\int _{0}^{1}dz \int _{0}^{1}\int _{0}
^{1}ds dt
\frac {z^{\epsilon   /2}}{t^{\epsilon   /2}} \frac {1}{(1-\frac 
{zt}{4})^{2+ 
\epsilon  
 /2}s^{1+\epsilon   }(1-s\tilde {\alpha})^{2+\epsilon   /2}},
\label{eq:3l11d}
\end{equation}
where $\tilde {\alpha} =\frac {(1-t/2)^2}{(1-zt/4)} $.
Consider the integration over $s$ first,
\begin{eqnarray}
\lefteqn{\int _{0}^{1}ds \frac {1}{s^{1+\epsilon   }(1-s\tilde
{\alpha})^{2+\epsilon   /2}}}\nonumber  \\
&=&\int _{0}^{1}ds \frac {s\tilde {\alpha}[2-s\tilde {\alpha}]
+[1-s\tilde {\alpha}]^2 \frac {\epsilon   }{2}\ln (1-s\tilde {\alpha
})+\cdot\cdot\cdot}{s^{1+\epsilon   }(1-s\tilde {\alpha })
^{2+\epsilon   /2}}
+\int _{0}^{1}ds \frac {1}{s^{1+ \epsilon  }}\nonumber  \\
&=&A_{1}+A_{2} .
\label{eq:3l11da}
\end{eqnarray}  
$A_1$ can be represented as the sum of  $A_{11}$ and $A_{12}$.
\begin{eqnarray}
A_{1}&=& A_{11}+A_{12}\nonumber  \\
&=&  \int _{0}^{1}ds \frac {\tilde {\alpha }}{s^{\epsilon   }(1-s\tilde
{\alpha})^{2+ \epsilon   /2}}+\int _{0}^{1}ds \frac {\tilde {\alpha
}}{s^{ \epsilon   }(1-s\tilde {\alpha})^{1+ \epsilon   /2}} 
\label{eq:3l11db}
\end{eqnarray}
Furthermore, the term $A_{11}$ can be decomposed into the
following:
\begin{eqnarray}
A_{11}&=&A_{111}+A_{112}\nonumber  \\
&=& \int _{0}^{1} ds \frac {\tilde {\alpha}}{(1-s\tilde
{\alpha})^{2+ \epsilon   /2}} +\epsilon   \int _{0}^{1}ds \frac {\ln s
\tilde {\alpha}}{(1-s\tilde {\alpha})^{1+ \epsilon   /2}} .
\end{eqnarray}

$A_{112}$ should not  concern us because it contains 
terms of at least first order in $\epsilon   $.   We only 
calculate $A_{111}$ as:
\begin{eqnarray}
A_{111}&=&[\frac {1}{(1+\epsilon   /2)(1-s\tilde {\alpha})^{1+ \epsilon  
/2}}]^{1}_{0}\nonumber  \\
&=& \frac {1}{(1+\epsilon   /2)}[-1+\frac {1}{(1-\tilde {\alpha})^{1+
\epsilon  
/2}}] .
\end{eqnarray}
Substituting $A_{111}$ into Eq.~(\ref{eq:3l11d}), we obtain:
\begin{eqnarray}
\lefteqn{ \int _{0}^{1}dz\int _{0}^{1}dt \frac {z^{\epsilon   /2}}
{(1+\epsilon 
/2)t^{\epsilon   /2}(1-\frac {zt}{4})^{2+ \epsilon 
 /2}}[\frac {1}{(1-\tilde
{\alpha})^{1+\epsilon   /2}}-1] }\nonumber  \\
&=& \int _{0}^{1}dz\int _{0}^{1}dt \frac {z^{\epsilon 
 /2}}{(1+\epsilon  
/2)t^{\epsilon   /2}(1-\frac {zt}{4})^{2+ \epsilon   /2}[1-\frac {(1-
t/2)^{2}}{(1-zt/4)}]^{1+ \epsilon   /2}}\nonumber  \\
&&-  \int _{0}^{1}dz\int _{0}^{1}dt \frac {z^{\epsilon   /2}}{t^{\epsilon 
/2}(1+\epsilon   /2)(1-\frac {zt}{4})^{2+ \epsilon   /2}}\nonumber  \\
&=&B_{1111}-B_{1112} ,
\label{eq:3l11dc}
\end{eqnarray}
where we separate the integrand into two parts:
\begin{eqnarray}
B_{1111}&=&\frac {1}{1+\epsilon   /2}\int _{0}^{1}\int _{0}^{1}dt
dz \frac {z^{\epsilon   /2}}{t^{\epsilon   /2}(1-\frac {zt}{4})t
^{1+\epsilon 
 /2}(1-
\frac {z}{4}-\frac {t}{4})^{1+\epsilon   /2}}\nonumber  \\
&=& \frac {1}{1+\epsilon   /2}\int _{0}^{1}
dz \quad z^{\epsilon   /2}   \int _{0}^{1}dt \frac {1}{(1-\frac
{zt}{4})t^{1+ \epsilon 
}(1-\frac {z}{4}-\frac {t}{4})^{1+\epsilon   /2}}
\label{eq:3l11e}
\end{eqnarray}
The integral over the variable $t$ in Eq.~(\ref{eq:3l11e}) is given by:
\begin{eqnarray}
\lefteqn{ \int _{0}^{1}dt \frac {z^{\epsilon   /2}}{t^{1+\epsilon  
 }(1-
\frac {z}{4}-\frac {t}{4})^{1+\epsilon 
 /2}}(\frac {1}{1-\frac {zt}{4}}-
1)+\int _{0}^{1}dt \frac {z^{\epsilon   /2}}{t^{1+\epsilon 
  }(1-\frac {z}{4}-
\frac {t}{4})^{1+\epsilon   /2}} }\nonumber  \\
&=& \int _{0}^{1}dt \frac {\frac {1}{4}z\times z^{\epsilon   /2}}
{t^{ \epsilon  
}(1-\frac {z}{4}-\frac {t}{4})^{1+\epsilon   /2}(1-\frac
{zt}{4})}
+ \int _{0}^{1}dt \frac {z^{\epsilon   /2}}{t^{1+\epsilon  }(1-\frac {z}{4}-
\frac {t}{4})^{1+\epsilon   /2}} \nonumber  \\
&=& C_{1111}+C_{1112}   .
\end{eqnarray}
The first term $C_{1111}$ is substituted into Eq.~(\ref{eq:3l11e}), and the
contribution is denoted by $B_{11111}$, which reads:
\begin{eqnarray}
B_{11111}&=&\int _{0}^{1} \int _{0}^{1}dt dz \frac {\frac
{1}{4}z}{(1-\frac {z}{4}-\frac {t}{4})(1-\frac {zt}{4})}
+ \cdots \mbox {irrelevant terms} \nonumber  \\
&=&4[9\ln (\frac {3}{4})+4 \ln 2-\frac {1}{2}(\ln 2)^{2}
+\Phi (1,2)-\Phi (\frac {1}{2},2)-\Xi(\frac {1}{4},2) ]
\end{eqnarray}
The evaluation of the integral is quite straightforward although
tedious. One can find the basic integrals in  Appendix C, whose
compositions will be used
to represent those complex integrals encountered in Eq.~(\ref{eq:3l11e}).

Here we take  the evaluation of $B_{11111}$ as an example and set
aside the rest of similar calculations.
\begin{eqnarray}
B_{11111} &=&\int _{0}^{1}\int _{0}^{1}dz dt \, \frac {1}{4}z [\frac
{1}{(1-\frac {z}{4}-\frac {t}{4})}-\frac {t}{(1-\frac {zt}{4})}]
\frac {1}{(1-t + \frac {t^2}{4})}\nonumber  \\
&=&\int _{0}^{1}\int _{0}^{1}dz dt \,
\frac {1}{\tilde {\Delta}}[\frac {(1-\frac {t}{4})}{(1-\frac
{z}{4}-\frac {t}{4})}-\frac {1}{(1-\frac {zt}{4})}]\nonumber  \\
&=&-4\! \int _{0}^{1}\! dt [\frac {(1-\frac {t}{4})}{(1-\frac {t}{2})^2
}\ln (\frac {3}{4}-\frac {t}{4})-\frac {\ln (1-\frac
{t}{4})}{(1-\frac {t}{2})^2
} (1-\frac {t}{4})-\frac {\ln (1-\frac {t}{4})}{(1-\frac
{t}{2})^2 t}], \nonumber \\
\label{eq:3l11f}
\end{eqnarray}
where $\tilde{\Delta}=1-t-\frac{t^2}{4}.$
Let $u=1-\frac {t}{2}$.   Eq.~(\ref{eq:3l11f}) turns into:
\begin{eqnarray}
Eq.~(\ref{eq:3l11f})&=&-2\times 1/2[\int _{1}^{1/2}du \frac {\ln (\frac
{1}{4}+\frac {u}{2})}{u^{2}}+\frac {\ln (\frac {1}{4}+\frac
{u}{2})}{u}-\frac {\ln (\frac {1}{2}+\frac {u}{2})}{u^{2}}- \nonumber \\
&& \frac {\ln (\frac {1}{2}+\frac {u}{2})}{u}
-\frac {\ln (\frac {1}{2}+\frac {u}{2})}{u^{2}}-
\frac {\ln (\frac {1}{2}+\frac {u}{2})}{u} +\frac {\ln (\frac {1}{2}
+\frac {u}{2})}{1-u}]  .
\label{eq:3l11g}
\end{eqnarray}
Each term in Eq.~(\ref{eq:3l11f})  can be easily represented in terms
of the basic integrals listed in  Appendix C.
For example,
\begin{eqnarray}
\int _{1}^{1/2}\frac {\ln (\frac {1}{4}+\frac {u}{2})}{u^2}
&=&2\ln (2)+3 \ln (\frac {3}{4}) \\
\int _{1}^{1/2} du \frac {\ln (\frac {1}{4}+\frac
{u}{2})}{u}
&=&[2\ln ^2(2)-\frac {1}{2}\ln ^2(2)+\Phi(\frac {1}{2},2)-
\Phi (1,2)]  \\
\int _{1}^{1/2}du [\frac {\ln (1+u)}{u^2}-
\frac {\ln (2)}{u^2}]
&=&-3\ln (\frac {3}{2})+2\ln (2) \\
\int _{1}^{1/2}du \frac {\ln (\frac {1}{2}+\frac {u}{2})}{u}
&=&\ln ^2(2)+\Phi(\frac {1}{2},2)-\Phi (1,2)\\
\int _{1}^{1/2}du \frac {\ln (\frac {1}{2}+\frac {u}{2})}{u^2}
&=&2\ln (2)-3\ln (\frac {3}{2})\\
\int _{1}^{1/2}du \frac {\ln (\frac {1}{2}+\frac {u}{2})}{(1-u)}
&=&\Xi (\frac {1}{4},2) \\
\int _{1}^{1/2}du \frac {(1+2u)}{u}
&=&-\frac {1}{2}\ln ^2 (2) -\Phi (1,2)+\Phi (\frac {1}{2},2)\\
\int _{1}^{1/2}du \frac {(1+u)}{u}
&=&\Phi(\frac {1}{2},2)-\Phi (1,2) ,
\end{eqnarray}
where $\Phi$ and $\Xi$ are defined in the Appendix C.

Now we turn to the calculation of $B_{11112}$, which is defined
as $B_{11112}=\int _{0}^{1}dt C_{1112}$.
\begin{eqnarray}
B_{11112}&=&\int _{0}^{1}\int _{0}^{1}dz dt \frac 
{z^{\epsilon   /2}}{t^{1+\epsilon 
 }
(1-\frac {z}{4}-\frac {t}{4})^{1+ \epsilon   /2}}\nonumber \\
&=&\int _{0}^{1}dz \frac {z^{\epsilon   /2}}{(1-\frac {z}{4})
^{1+ \epsilon  /2}}
\int _{0}^{1}dt \frac {1}{t^{1+\epsilon   }[1-\frac {t}{4(1-z/4)}]
^{1+\epsilon 
 /2}} ,
\label{eq:3l11j}
\end{eqnarray}
where  the $t$-dependent integral can be separated into:
\begin{eqnarray}
\lefteqn{\int _{0}^{1}dt \frac {1}{t^{1+\epsilon   }(1-\tilde 
{\beta}t)^{1+
\epsilon   /2}}}\nonumber  \\
&=&\int _{0}^{1}dt\frac {1}{t^{1+ \epsilon  }}\frac {[1-(1-\tilde
{\beta}t)^{1+\epsilon   /2}]}{(1-\tilde {\beta})^{1+ \epsilon  /2}}
+\int _{0}^{1}dt \frac {1}{t^{1+\epsilon   }}\nonumber  \\
&=&\int _{0}^{1}dt \frac {1}{t^{\epsilon   }}\frac {\tilde {\beta}}{(1-
\tilde {\beta})^{1+\epsilon   /2}} +O(\epsilon   )+\int _{0}^{1}dt \frac
{1}{t^{1+\epsilon   }}\nonumber  \\
&=& D_1+D_2+O(\epsilon   )
\label{eq:3l11h}
\end{eqnarray}
with $\tilde {\beta}$ being $\frac {1}{4-z}$.
The term $D_1$ in Eq.~(\ref{eq:3l11h}) is
\begin{equation}
D_1=\int _{0}^{1}dt \frac {\tilde {\beta}}{(1-\tilde {\beta}t)}=
-\ln (1-\tilde {\beta}).
\end{equation}
After substituting the above equation into Eq.~(\ref{eq:3l11j}), one has:
\begin{eqnarray}
\int _{0}^{1}dz  (-1) \frac {z^{\epsilon   /2}\ln (1-\frac {1}{4-z})}
{(1-\frac {z}{4})^{1+\epsilon   /2}}
&=&\int _{0}^{1}dz \frac {[\ln (4-z)-\ln (3-z)]}{(1-
\frac {z}{4})}+O(\epsilon   )\nonumber  \\
&=&I_A-I_B+4\ln ^2(\frac {4}{3}) .
\end{eqnarray}

Inserting  $D_2$ in Eq.~(\ref{eq:3l11h}) into Eq.~(\ref{eq:3l11j}), 
we have:
\begin{eqnarray}
 &=&\int _{0}^{1}dt \frac {1}{t^{1+\epsilon   }}\times
\int _{0}^{1}dz \frac {z^{\epsilon   /2}}{(1-\frac {z}{4})
^{1+\epsilon   /2}}
\nonumber  \\ &=& \frac {-1}{\epsilon  }\times [\int _{0}^{1}dz
\frac {1}{(1-\frac {z}{4})}-\frac {\epsilon   }{2}\int _{0}^{1}dz
\frac {\ln (1-\frac {z}{4})}{(1-\frac {z}{4})}+
\int _{0}^{1}dz \frac {\ln (z)}{(1-\frac {z}{4})}]+O(\epsilon   )
\nonumber  \\&=&(-\frac {1}{\epsilon   })[4\ln (\frac {4}{3})
-\frac {\epsilon  }{2}I_A+\frac {\epsilon   }{2}I_C]
\end{eqnarray}

Now we move on to calculate the 
contribution of $A_2$ in Eq.~(\ref{eq:3l11da}).
By inserting $A_2$ into Eq.~(\ref{eq:3l11d}), one has:
\begin{eqnarray}
\lefteqn{\int _{0}^{1}\int _{0}^{1}\int _{0}^{1}dz dt ds
\frac {z^{\epsilon  /2}}{t^{\epsilon   /2}}\frac {1}{s^{1+\epsilon 
 }(1-\frac
{zt}{4})^{2+\epsilon   /2}}} \nonumber  \\
&=& \int _{0}^{1}  \int _{0}^{1}\int _{0}^{1}dz ds dt
\frac {1}{s^{1+\epsilon   }(1-\frac {zt}{4})^{2}}+
\int _{0}^{1}\int _{0}^{1}\int _{0}^{1}dz dt ds \nonumber  \\
&& \frac {1}{s^{1+\epsilon   }(1-\frac {zt}{4})^{2}}
[(-\frac {\epsilon   }{2})
\ln (1-\frac {zt}{4})+\frac {\epsilon   }{2}\ln (z)-\frac 
{\epsilon   }{2}
\ln (t)]+ \cdots\nonumber  \\
&=&4\ln (\frac {4}{3})\frac {1}{-\epsilon   }+\int _{0}^{1}\int _{0}^{1}
dz dt (\frac {1}{-\epsilon   })(\frac {\epsilon   }{2})\nonumber  \\
&& [\frac {-\ln (1-\frac {zt}{4})+\ln (z)-\ln (t)}{(1-\frac
{zt}{4})^{2}}] .
\label{eq:3l11k}
\end{eqnarray}
Here we decompose the second term in 
Eq.~(\ref{eq:3l11k}) into three parts, $P_1$ , $P_2$,
and $P_3$.

\begin{eqnarray}
P_1&=&\int _{0}^{1}\int _{0}^{1}dz dt\frac {\ln (1-\frac
{zt}{4})}{(1-\frac {zt}{4})^{2}}\nonumber \\
&=&\int _{0}^{1}dt
 (\frac {-4}{t}) \frac {\ln (1-\frac {zt}{4})}{(1-\frac
{zt}{4})}|^{1}_{0}+\int _{0}^{1}\int _{0}^{1}dz dt \frac {(-1)}
{(1-\frac
{zt}{4})^2}\nonumber  \\
&=& \int _{0}^{1}dt\quad \frac {(-4)}{t}\frac {\ln (1-\frac
{t}{4})}
{(1-\frac {t}{4})}+(-4\ln (\frac {4}{3}))\nonumber  \\
&=&-\int _{0}^{1}dt \quad 4\frac {\ln (1-\frac {t}{4})}{t}
-\int _{0}^{1}dt \frac {\ln (1-\frac {t}{4})}{(1-\frac {t}{4})}
-4 \ln (\frac {4}{3})\nonumber  \\
&=&-[4I_D+I_A]-4\ln (\frac {4}{3})\\
P_2&=&\int _{0}^{1}\int _{0}^{1}dt dz \frac {\ln (z)}{(1-\frac
{zt}{4})^2}    \nonumber \\
&=&\int _{0}^{1}\frac {\ln (z)}{(1-\frac {z}{4})}
= I_C  \\
P_3&=&\int _{0}^{1}\int _{0}^{1}dt dz \frac {\ln (t)}{(1-\frac
{zt}{4})^2}=I_C.
\end{eqnarray}

Now  we  go back to finish the calculation
of $A_{12}$ in Eq.~(\ref{eq:3l11db}).
\begin{eqnarray}
A_{12}&=&\int _{0}^{1} \int _{0}^{1}dzdt \frac {\ln (1-\frac {(1-
t/2)^2}{1-\frac {zt}{4}})}{(1-\frac {zt}{4})^2} \nonumber  \\
&=&-\int _{0}^{1} \int _{0}^{1}dz dt \frac {\ln (1-\frac
{zt}{4})}{(1-\frac {zt}{4})^2}
+\int _{0}^{1}\int _{0}^{1}dz dt\frac {\ln (t) +\ln (1-\frac
{z}{4}-\frac {t}{4})}{(1-\frac {zt}{4})^{ 2}}\nonumber  \\
&=& [4I_D+I_A+4\ln (\frac {4}{3})]+I_C+\int _{0}^{1}dt
\frac {4}{t}\frac {\ln (1-\frac {z}{t}-\frac {t}{4})}
{(1-\frac {zt}{4})}|^{1}_{0}\nonumber  \\
&&-\int _{0}^{1}\int _{0}^{1}dz dt\frac {4\times (-\frac
{1}{4})}{t(1-\frac {zt}{4})(1-\frac {z}{4}-\frac {t}{4})}
\nonumber  \\ &=&[4I_D+I_A+4\ln (\frac {4}{3})]+I_C+ L_1 -M_1+M_2 ,
\label{eq:3l11m}
\end{eqnarray}
where $L_1$, $M_1$ and $M_2$  are analyzed below.
\begin{eqnarray}
L_1&=&\int _{0}^{1}dt \frac {4}{t}[\frac {\ln (\frac {3}{4}-\frac
{t}{4})}{(1-\frac {t}{4})}-\ln (1-\frac {t}{4})]\nonumber  \\
&=&\int _{0}^{1}dt \{ [\frac {4}{t}+\frac {1}{(1-\frac {t}{4})}]
\ln (\frac {3}{4}-\frac {t}{4})-\frac {4}{t}\ln (1-\frac
{t}{4})\}
\nonumber  \\
&=& \int _{0}^{1}dt \frac {4}{t}\ln (\frac {3}{4}-\frac {t}{4})
+[-4 \ln (\frac {3}{4})+I_B-4I_D] .
\label{eq:3l11n}
\end{eqnarray}
The first term in Eq.~(\ref{eq:3l11n}) is divergent.
As we will show later,  it will be canceled by a
term  in $M_2$.
\begin{eqnarray}
M_1&=&8\int _{0}^{1}dt \frac {1}{t}(\frac {-1}{\Delta}+1)\ln (1-
\frac {t}{4})-8\int _{0}^{1}dt \frac {1}{t}\ln (1-\frac {t}{4})
\nonumber  \\ &=& 8\int _{1}^{1/2}du[\frac {\ln (\frac
{1}{2}+\frac {u}{2})}{u^2}+\frac {\ln (\frac {1}{2}+\frac
{u}{2})}{u}]-8\int _{0}^{1}dt \frac {1}{t}\ln (1- \frac {t}{4})
\nonumber  \\&=&
8[2\ln (2)-3\ln (\frac {3}{2})+\ln ^2(2)+\Phi (\frac {1}{2},2)-
\Phi (1,2)] -8I_D  ,
\end{eqnarray}
where $\Delta=(1-\frac {t}{2})^{2}$ and $ u=1-\frac {t}{2}$.
\begin{eqnarray}
M_2&=&\int _{0}^{1}dt \frac {4}{t}\ln (\frac {3}{4}-\frac
{1}{4}t)(\frac {-1}{\Delta}+1)-\int _{0}^{1}dt\quad \frac {4}{t}
\quad \ln (\frac {3}{4}-\frac {t}{4})\nonumber  \\
&=& 4\int _{1}^{1/2}du (\frac {1+u}{u^2})\ln (\frac {1}{4}+\frac
{u}{2})-\int _{0}^{1}dt \frac {4}{t}\ln (\frac {3}{4}-\frac
{t}{4})\nonumber  \\
&=&4[2\ln (2)+3\ln (\frac {3}{4})+2\ln ^2(2)-\frac {1}{2}
\ln ^2(2)+\Phi (\frac {1}{2},2)-\Phi (1,2)]\nonumber  \\
&&-\int _{0}^{1}dt \frac {4}{t}\ln (\frac {3}{4}-\frac {t}{4}) ,
\label{eq:3l11o}
\end{eqnarray}
where, as mentioned above, the first term  in Eq.~(\ref{eq:3l11n})
is canceled by the last term in Eq.~(\ref{eq:3l11o}).

The term $B_{1112}$ in Eq.~(\ref{eq:3l11dc}) is calculated below.
\begin{eqnarray}
B_{1112}&=&\frac {1}{1+\epsilon   /2}\int _{0}^{1}\int _{0}^{1}dz dt
\frac {z^{\epsilon   /2}}{t^{\epsilon   /2}(1-\frac {zt}{4})^{2+\epsilon 
 /2}}\nonumber
\\ &=& \int _{0}^{1}\int _{0}^{1}dz dt \frac {1}{(1-\frac
{zt}{4})^{ 2}} +O(\epsilon   )\nonumber  \\
&=&4\ln (\frac {4}{3})+O(\epsilon 
 ).
\end{eqnarray}

The second term $3.1-2$ in Eq.~(\ref{eq:3l1a}) is calculated as follows:
\begin{eqnarray}
\lefteqn {3.1-2}\nonumber  \\
&=&\int _{-\infty}^{\infty}\! d^d \vec p\int _{-\infty}^{\infty}\! d^d
 \vec q\int _{-\infty}^{\infty}\! d^d \vec k \frac {\vec p\cdot\vec
 q }{(\! q^2 \!+\! m^2\! )[(\!\vec p\!-\!  \vec k\!)^2 \!+\! m^2]
\{\! (\!k^2
\!+\!  m^2\!)
 \!+\! [k^2\!+\! (\! \vec p\!-\! \vec k\!)^2\!+\! 2m^2]\! \}}\nonumber  \\
&&\frac {1}{\{[q^2+(\vec p -\vec q)^2+2m^2]+[k^2+(\vec p-\vec
 k)^2+2m^2]\} } \nonumber  \\
&=&\frac {\Gamma (1+1+1)}{4[\Gamma  (1)]^3}\int _{-
 \infty}^{\infty} d^d \vec p\int _{-\infty}^{\infty} d^d \vec
 q\int _{-\infty}^{\infty} d^d \vec k
 \frac {\vec p\cdot\vec q }
 {(q^2+m^2)} {\int _{0}^{1}\int _{0}^{1}}_{x+y<1}dx dy \nonumber  \\
&&\{(1-x-y)[(\vec p-\vec k)^2+m^2]
 +x(k^2+p^2-\vec k\cdot\vec p+m^2) \nonumber  \\
&&+y(p^2+q^2+k^2
 -\vec p\cdot\vec q -\vec p\cdot\vec k+m^2)\}^{-3}+\mbox
 {irrelevant terms} \nonumber  \\
&=& \frac{\Gamma (3)}{4} \int_{-\infty}^{\infty} d^d \vec p
 \int _{-\infty}^{\infty} d^d \vec q \frac {\vec p\cdot\vec q}
 {(q^2+m^2)}{\int _{0}^{1}\int _{0}^{1}}_{x+y<1}dx dy \int_{0}^{1}
 dz (\pi)^{d/2} \nonumber  \\
&&\frac {\Gamma (3-1-\epsilon 
 /2)}{\Gamma (3)} \frac {1}
 {\{[1-(1-\frac {x+y}{2})^2]p^2+\vec p\cdot\vec q (-y)
 +yq^2+m^2\}^{2-\epsilon 
 /2}}\nonumber  \\
&=&(\pi)^{d/2}\frac {\Gamma (2-\epsilon 
 /2)}{4}\int _{-\infty}^{\infty}
 d^d \vec q {\int _{0}^{1}\int _{0}^{1}}_{x+y<1}dx dy
 \frac {1}{(q^2+m^2)}(\pi)^{d/2} \frac {\Gamma (1-\epsilon 
 )}{\Gamma (2-
 \epsilon   /2)} \nonumber \\
&&\frac {\vec q \cdot (\frac {y}{2\Delta} \vec q)}
 {(\Delta)^{2-\epsilon 
 /2}\{[\frac {y}{\Delta}-(\frac
 {y}{2\Delta})^2]q^2+\frac {m^2}{\Delta}\}^{1-\epsilon 
 }}\nonumber  \\
&=&\pi^{d}\frac {\Gamma (1-\epsilon 
 )}{8}{\int _{0}^{1}\int_{0}^{1}}_{x+y<1}
 dx dy \frac {y}{\Delta ^{3-\epsilon 
 /2}\Box ^{1-\epsilon 
}}
 [\frac {m}{(\Box \Delta)^{1/2}}]^{2+\epsilon 
 -2(1- \epsilon 
)}\pi ^{d/2}
 \frac {\Gamma (-\frac {3}{2}\epsilon 
 )}{\Gamma (1-\epsilon 
)}\nonumber  \\
&=&\frac {\pi ^{3d/2}\Gamma (-\frac {3}{2}\epsilon 
 )}{8}
 (m^2)^{3/2 \epsilon 
 }{\int _{0}^{1}\int _{0}^{1}}_{x+y<1}dx dy
 \frac {y}{\Delta ^{3+\epsilon 
 }(\frac {y}{\Delta}-\frac {y^2}{4\Delta
 ^2})^{1+\epsilon 
 /2}},\nonumber \\ 
\label{eq:3l12i}
\end{eqnarray}
where $\Delta =(x+y)^2 -\frac {1}{4}(x+y)^2,
\Box =\frac {y}{\Delta}-\frac {y}{2\Delta} $.

The integral  in Eq.~(\ref{eq:3l12i}) is analyzed below.
\begin{eqnarray}
 \lefteqn{{\int _{0}^{1}\int _{0}^{1}}_{x+y<1}dx dy
 \frac {y}{\Delta ^{3+\epsilon 
 }} \frac {1}{[\frac {y^{1+\epsilon 
 /2}}{\Delta
 ^{2+\epsilon 
 }}](\Delta-\frac {y}{4})^{1+\epsilon 
 /2}}}\nonumber  \\
&=&\int _{0}^{1}\int _{0}^{1}ds dt \frac {s\times st}{s(1-\frac
 {s}{4})s^{1+\epsilon 
 /2}t^{1+\epsilon   /2}}\frac {1}{(s-\frac {s^2}{4}-
 \frac {st}{4})^{1+\epsilon   /2}}\nonumber  \\
&=&\int _{0}^{1}\int _{0}^{1}ds dt \frac {1}{s^{1+\epsilon 
 }t^{\epsilon  /2 }
 (1-\frac {s}{4}-\frac {t}{4})^{1+\epsilon 
 /2}}[\frac {1}{1-\frac
 {s}{4}}-1]      \nonumber \\
&&+\int _{0}^{1}\int _{0}^{1}ds dt
 \frac {1}{s^{1+\epsilon   }t^{\epsilon   /2}}\frac {1}{(1-\frac 
{s}{4}-\frac
 {t}{4})^{1+\epsilon   /2}}\nonumber  \\
&=&S+R_1+R_2+O(\epsilon   ) ,
\end{eqnarray}
where  the transformations of  $y=st$ and  $x=s(1-t)$ have been used,
and the calculations  for $S$, $R_1$ and $R_2$ will be 
presented successively.

\begin{eqnarray}
S&=&\int _{0}^{1}\int _{0}^{1}ds dt \frac {\frac {1}{4}}{(1-\frac
{s}{4}-\frac {t}{4})(1- \frac {s}{4})} \nonumber \\
&=&\int _{0}^{1} ds\frac {1}{1-\frac {s}{4}}
[-\ln (\frac {3}{4}-\frac {s}{4})
+\ln (1-\frac {s}{4})]  \nonumber \\
&=&[I_A+4\ln ^2(\frac {4}{3})-I_B]
\end{eqnarray}

The reason for the rise of $R_1$ and $R_2$ can be revealed by the
following decomposition:
\begin{eqnarray}
R_1 + R_2 &=&\int _{0}^{1}\int _{0}^{1}ds dt \frac {1}{s^{1+\epsilon 
 }t^{\epsilon   /2}
[1-\frac {s}{4}-\frac {t}{4}]^{1+\epsilon   /2}}\nonumber \\
&=&\int _{0}^{1}\int_{0}^{1}ds dt  \frac {1}{t^{\epsilon   /2}(1-\frac
 {t}{4})^{1+\epsilon   /2}}\frac {1}{s^{1+\epsilon   }(1-\frac
{s}{4-t})^{1+\epsilon   /2}}   \nonumber \\
&=&\int _{0}^{1}dt \frac {1}{t^{\epsilon   /2}(1-\frac
{t}{4})^{1+\epsilon   /2}}\int _{0}^{1}ds \frac {1}{s^{1+\epsilon   }}
[\frac {1}{(1-\frac {s}{4-t})^{1+\epsilon    /2}}-1] \nonumber \\
&&+\int_{0}^{1}dt \frac {1}{t^{\epsilon  /2}(1-\frac {t}{4})^{1+\epsilon 
  /2}}
\int _{0}^{1} ds \frac  {1}{s^{1+\epsilon   }} ,
\end{eqnarray}
where
\begin{eqnarray}
R_1&=&\int _{0}^{1}dt \frac {1}{t^{\epsilon /2}(1-\frac {t}{4})
^{1+\epsilon
/2}}\int _{0}^{1}ds \frac {1}{s^{\epsilon }(4-t)(1-\frac {s}{4-
t})^{1+\epsilon /2}}\nonumber  \\
&=&\int _{0}^{1}dt \frac {1}{(1-\frac {t}{4})(4-t)}\int _{0}^{1}ds
\frac {1}{(1-\frac {s}{4-t})} +O(\epsilon )\nonumber  \\
&=&\int _{0}^{1}dt \frac {[\ln (4-t)-\ln (3-t)]}{(1-\frac
{t}{4})}=4\ln ^2(\frac {4}{3})+I_A-I_B  ,
\end{eqnarray}
and
\begin{eqnarray}
R_2 &=&\int _{0}^{1}dt \frac {1}{t^{\epsilon }(1-\frac {t}{4})^{1+
\epsilon
/2}}\int _{0}^{1}ds \frac {1}{s^{1+\epsilon }}\nonumber \\
&=&\int _{0}^{1}dt \frac {1}{1-\frac {t}{4}}(\frac {1}{-\epsilon })
+ \int _{0}^{1}dt \frac {[\ln (t)+1/2\ln (1-\frac {t}{4})]}
{(1-\frac {t}{4})} +O(\epsilon )\nonumber  \\
&=& 4\ln (\frac {4}{3})(\frac {1}{-\epsilon })+I_C+\frac {1}{2}I_A  .
\end{eqnarray}

The term $3-1.3$ will be equal to $3-2.2$, which will be shown in
next two subsections.

For the other 3-loop diagram, we can still employ the same technique.
To begin with, we write down the corresponding integral for Fig. 15.
\begin{eqnarray}
\mbox {FD 3l2}=&&\int _{-\infty}^{\infty} d^d \vec p\int _{-\infty}
^{\infty} d^d
\vec q\int _{-\infty}^{\infty} d^d \vec k \frac {(-\vec
p\cdot\vec k)\times (-\vec p\cdot\vec q
)}{(k^2+m^2)(p^2+m^2)(q^2+m^2)}
\int _{0}^{\infty}dt_x \int _{-\infty}^{t_x}dt_y \nonumber  \\
&& e^{-[k^2+m^2+(\vec p-\vec k)^2+m^2](t_x-t_y)}e^{-
[q^2+m^2+(\vec p -\vec q)^2+m^2]t_x}e^{-(p^2+m^2)|t_y|}.
\label{eq:3l2i}
\end{eqnarray}
Again, the time dependent part of the integral in
Eq.~(\ref{eq:3l2i}) can be integrated
first.
For convenience, let $a=(p^2+m^2), b=[(\vec p -\vec q)^2+m^2], 
c=(k^2+m^2),
d=[(\vec p-\vec k)^2+m^2],$ and $e=(p^2+m^2)$.
 Then we have:
\begin{eqnarray}
\lefteqn{\int _{0}^{\infty}dt_x e^{-[c+d+a+b]t_x}[\int _{0}^{t_x}
dt_y e^{(c+d-e)t_y}+\int _{-\infty}^{0}dt_y e^{(c+d+e)t_y}]}
\nonumber  \\
&=& \int _{0}^{\infty}dt_x e^{-(a+b+c+d)t_x}
[\frac {1}{c+d-e}(e^{(c+d-e)t_x}-1)+\frac {1}{c+d+e}] \nonumber \\
&=&\frac {2}{(c+d+e)(a+b+c+d)}   .
\end{eqnarray}

In the same spirit of the calculation as  performed in
the previous case, we separate the integrals into
several pieces, each of which  only  contains an isolated  pole,
and then extract the corresponding singular parts.

One can start with decomposing $\vec p\cdot\vec k$ into:
\begin{equation}
\vec p\cdot\vec k=\frac {1}{2}[-(p^2+k^2+(\vec p-\vec k)^2+3m^2
)+2(p^2+m^2)+2(k^2+m^2)] .
\end{equation}
Then one can rewrite Eq.~(\ref{eq:3l2i}) as $[-(3.2-1)+2(3.2-2)+2(3.2-3)]$,
where
\begin{eqnarray}
3.2-1 &=&\int _{-\infty}^{\infty} d^d \vec p
\int _{-\infty}^{\infty} d^d \vec q\int _{-\infty}^{\infty} d^d
\vec k \frac {\vec p\cdot\vec q }{(k^2+m^2)(q^2+m^2)(p^2+m^2)}
\nonumber  \\
&& \frac {1}{[k^2+q^2+(\vec p -\vec q)^2+(\vec p-
\vec k)^2+4m^2]}\\
3.2-2&=&\int _{-\infty}^{\infty} d^d \vec p
\int _{-\infty}^{\infty} d^d \vec q\int _{-\infty}^{\infty} d^d
\vec k \frac {\vec p\cdot\vec q
}{(k^2+m^2)(q^2+m^2)[p^2+k^2+(\vec p-\vec k)^2+3m^2]}
\nonumber  \\
&& \frac {1}{[k^2+q^2+(\vec p -\vec q)^2+(\vec p -\vec k)^2+4m^2]}\\
3.2-3&=&\int _{-\infty}^{\infty} d^d \vec p
\int _{-\infty}^{\infty} d^d \vec q\int _{-\infty}^{\infty} d^d
\vec k \frac {\vec p\cdot\vec q
}{(p^2+m^2)(q^2+m^2)[p^2+k^2+(\vec p-\vec k)^2+3m^2]}
\nonumber  \\
&& \frac {1}{[k^2+q^2+(\vec p -\vec q)^2+(\vec p
-\vec k)^2+4m^2]}.
\end{eqnarray}
Again, we have three different types of the integrals to handle.

The calculation for the term $3.2-1$ is shown as below.
\begin{eqnarray}
\lefteqn{3.2-1}\nonumber \\
&=&\int _{-\infty}^{\infty} d^d \vec p
 \int _{-\infty}^{\infty} d^d \vec q\int _{-\infty}^{\infty} d^d
 \vec k \frac {\vec p\cdot\vec q }{(k^2+m^2)(q^2+m^2)(p^2+m^2)}
 \times \nonumber  \\
&& \frac {1}{k^2+q^2+(\vec q -\vec p)^2+(\vec p-\vec k)^2+4m^2}
\nonumber \\
&=&\frac {1}{2}\int _{-\infty}^{\infty} d^d \vec p
 \int _{-\infty}^{\infty} d^d \vec q \frac {\vec p\cdot\vec q }
 {(p^2+m^2)(q^2+m^2)}\int _{0}^{1}dx \int _{-\infty}^{\infty} d^d
 \vec k \nonumber  \\
&& \frac {1}{[k^2-x\vec p\cdot\vec k+xp^2+xq^2
 -x\vec p\cdot\vec q +m^2]^2}\nonumber  \\
&=&\frac {1}{2}\int _{-\infty}^{\infty} d^d \vec q
 \frac {\vec p\cdot\vec q }{(p^2+m^2)(q^2+m^2)}\nonumber \\
&&\int _{0}^{1}dx\pi ^{d/2}\frac {\Gamma (1-\epsilon /2)}{\Gamma (2)}
 \frac {1}{[(x-\frac {x^2}{4})p^2-x\vec p
 \cdot\vec q +xq^2+m^2]^{1-\epsilon /2}}\nonumber  \\
&=&\frac {\pi ^{d/2}\Gamma (1-\epsilon /2)}{2}
 \int _{-\infty}^{\infty} d^d \vec p \int _{-\infty}^{\infty} d^d
 \vec q \frac {\vec p\cdot\vec q }{(p^2+m^2)(q^2+m^2)}\nonumber\\
&&\frac {1}{(x-\frac {x^2}{4})^{1-\epsilon /2}(p^2-\frac{x}{\Delta}
 \vec p\cdot\vec q +\frac {x}{\Delta}q^2+\frac {m^2}{\Delta})
 ^{1-\epsilon /2}}\nonumber  \\
&=& \frac {\pi ^{d/2}\Gamma (1-\epsilon /2)}{2} \frac {\Gamma (2-\epsilon
 /2)}{\Gamma (1-\epsilon /2)}\int _{-\infty}^{\infty} d^d \vec p
 \int _{-\infty}^{\infty} d^d \vec q
 \int _{0}^{1}\int _{0}^{1}dx dy
 \frac {1}{(q^2+m^2)\Delta ^{1-\epsilon /2}}\nonumber  \\
&&\frac {\vec p\cdot\vec q
 y^{-\epsilon /2}}{[p^2-\frac {xy}{\Delta}\vec p\cdot\vec q +
 \frac {xy}{\Delta}q^2 +\frac {y}{\Delta}m^2]^{2-\epsilon /2}}
 \nonumber  \\
&=&\frac {\pi ^{d}\Gamma (1-\epsilon )}{2} \int _{0}^{1}
 \int _{0}^{1}dx dy \int _{-\infty}^{\infty} d^d \vec q
 \frac {y^{-\epsilon /2}}{(q^2+m^2)\Delta ^{1-\epsilon /2}} \nonumber \\
&& \frac {\vec q
 \cdot(\frac {xy}{2\Delta}\vec q)}{\{[\frac {xy}{\Delta}-(
 \frac {xy}{2\Delta})^{2}]q^2+\frac {y}{\Delta}m^2\}^{1-\epsilon}}
 \nonumber  \\
&=&\frac {\pi ^d\Gamma (1-\epsilon )}{2}
 \int _{0}^{1}\int _{0}^{1}dx dy \int _{-\infty}^{\infty} d^d \vec
 q \frac {(xy)y^{-\epsilon /2}}{2\Delta \Box ^{1-\epsilon }\Delta 
^{1-\epsilon /2}}
 \frac {1}{(q^2+\frac {y}{\Box \Delta}m^2)^{1-\epsilon }}\nonumber  \\
&=&\frac {\pi ^{3d/2}}{4}\Gamma (-\frac {3\epsilon   }{2})(m^2)^{3\epsilon 
 /2}
\int _{0}^{1}\int _{0}^{1}dx dy \frac {xy^{1+3\epsilon   /2}y^{-\epsilon /2}}
{\Box ^{1+\epsilon /2}\Delta ^{2+\epsilon }} ,
\label{eq:3l21i}
\end{eqnarray}
where  $\Delta =x -\frac {x^2}{4}$ and $ \Box =\frac {xy}{\Delta}
-(\frac {xy}{2\Delta})^2$.
The integration over $x$ and $y$ in  Eq.~(\ref{eq:3l21i}) is carried out
as below.
\begin{eqnarray}
\lefteqn{ \int _{0}^{1}\int _{0}^{1} dx dy \frac {xy^{1+\epsilon }}
{(xy)^{1+\epsilon /2}(\Delta -\frac {xy}{4})^{1+\epsilon /2}}}
\nonumber  \\
&=& \int _{0}^{1}\int _{0}^{1} dx dy \frac {y^{\epsilon /2}}{x^
{1+\epsilon }
(1-\frac {x}{4}-\frac {y}{4})^{1+\epsilon /2}}\nonumber  \\
&=&\int _{0}^{1} dy\frac {y^{\epsilon /2}}{(1-\frac {y}{4})^{1+
\epsilon /2}}
\int _{0}^{1}dx \frac {1}{x^{1+\epsilon }(1-\frac {x}{4-y})^{1+ 
\epsilon/2 }}.
\label{eq:3l21a}
\end{eqnarray}
Rewrite Eq.~(\ref{eq:3l21a}) as:
\begin{eqnarray}
\mbox{Eq}.~(\ref{eq:3l21a})&=& \int _{0}^{1}dy \frac {y^{\epsilon /2}}
{(1-\frac {y}{4})^{1+\epsilon
/2}}[\int _{0}^{1}dx \frac {1}{x^{1+\epsilon }} [\frac {1}{(1-\frac
{x}{4-y})^{1+ \epsilon/2}} -1]+\int _{0}^{1} \frac {1}{x^{1+\epsilon
}}]\nonumber  \\
&=&\int _{0}^{1}dy \frac {y^{\epsilon /2}}{(1-\frac {y}{4})^{1+\epsilon /2}}
[-\ln (1-\frac {x}{4-y})|^{1}_{0}+\frac {1}{-\epsilon }+O(\epsilon )].
\label{eq:3l21b}
\end{eqnarray}
One also  expands the terms in Eq.~(\ref{eq:3l21a})
 up to the zero order in $\epsilon $.
The finite part of the first term in Eq.~(\ref{eq:3l21b})  is found to be:
\begin{equation}  
\int _{0}^{1}dy \frac {1}{(1-\frac {y}{4})}[\ln (4-y)-\ln (3-y)]
= 4 (\ln \frac {4}{3})^2 +I_A -I_B  .
\end{equation} 
The second part  is extracted in the same manner:
\begin{eqnarray}
\int _{0}^{1}dy \frac {y^{\epsilon /2}}{(1-\frac {x}{4-y})^{1+\epsilon /2
}}
&=&\{\int _{0}^{1}dy \frac {1}{(1-\frac {y}{4})}+ \frac {\epsilon }{2}
[\frac {\ln (y)}{(1-\frac {y}{4})}-\frac {\ln (1-\frac
{y}{4})}{1-\frac {y}{4}}]\}+O(\epsilon )\nonumber  \\
&=& [4\ln (\frac {4}{3})+\frac {\epsilon }{2}(I_C - I_A)].
\end{eqnarray}

Now we turn to the term $3.2-2$.
\begin{eqnarray}
\lefteqn{\mbox {3.2-2}}\nonumber \\
&=&\int _{-\infty}^{\infty} d^d \vec p\int _{-\infty}^{\infty} d^d
 \vec q\int _{-\infty}^{\infty} d^d \vec k
 {\int _{0}^{1}\int _{0}^{1}}_{x+y<1}dx dy \frac {1}{4(q^2+m^2)}
 \frac {\vec p\cdot\vec q }{(q^2+m^2)}\nonumber  \\
&& \frac {1}{[k^2-(x+y)\vec
 p\cdot\vec k+(x+y)p^2+yq^2-y\vec p\cdot\vec q +m^2]^{3}}\nonumber  \\
&=&\frac {\Gamma (3)}{(\Gamma (1)^3}\int _{-\infty}^{\infty} d^d
 \vec p\int _{-\infty}^{\infty} d^d \vec q \frac {1}{4(q^2+m^2)}
 {\int _{0}^{1}\int _{0}^{1}}_{x+y<1}dx dy \pi^{d/2}
 \frac {\Gamma (2-\epsilon /2)}{\Gamma (3)}\nonumber  \\
&& \frac {\vec p\cdot\vec q }
 {\Delta ^{2-\epsilon /2}[p^2+\frac {y}{\Delta}q^2-\frac {y}{\Delta}
 \vec p\cdot\vec q +\frac {m^2}{\Delta}]^{2-\epsilon /2}}\nonumber  \\
&=&\int _{-\infty}^{\infty} d^d \vec q
 \frac {1}{4(q^2+m^2)}{\int _{0}^{1}\int _{0}^{1}}_{x+y<1}dx dy
 \pi ^{d}\frac {\Gamma (2-\epsilon /2)}{\Gamma (3)}\frac {\Gamma 
(1-\epsilon
 )}{\Gamma (2-\epsilon /2)} \nonumber  \\
&& \frac {\vec q\cdot(\frac {y}{2\Delta})\vec q}
 {\Delta ^{2-\epsilon /2}\{[\frac {y}{\Delta}-(\frac
 {y}{2\Delta})^2]q^2+\frac {m^2}{\Delta}\}^{1-\epsilon }}\nonumber  \\
&=&\pi^{d}\frac {\Gamma (1-\epsilon )}{8}\int _{-\infty}^{\infty} 
d^d \vec
 q \frac {y}{\Delta ^{3-\epsilon /2}\Box ^{1-\epsilon }} \frac
 {q^2}{(q^2+\frac {m^2}{\Box \Delta})^{1-\epsilon }(q^2+m^2)} 
\nonumber \\
&& (\pi)^{d/2}\frac {\Gamma (2+\epsilon /2)}
 {\Gamma (1+\epsilon /2)} \frac {\Gamma ( -3\epsilon   /2)}{\Gamma 
(1-\epsilon)}
 \nonumber  \\
&=& \pi ^{3d/2} \Gamma (-3\epsilon   /2)
 {\int _{0}^{1}\int _{0}^{1}}_{x+y<1}dx dy \frac {1}{8}\frac
 {\Gamma (2+\epsilon /2)}{\Gamma (1+\epsilon /2)}\frac {y}{\Delta 
^{3+\epsilon }
 \Box ^{1+\epsilon /2}} .
\label{eq:3l22i}
\end{eqnarray}
Here $\Delta = (x+y)-\frac {1}{4}(x+y)^2 $ and
$ \Box =\frac {y}{\Delta}-(\frac {y}{2\Delta})^2$.

The integration over $x$ and $y$ in Eq.~(\ref{eq:3l22i})  is performed
as below.
\begin{eqnarray}
&&{\int _{0}^{1}\int _{0}^{1}}_{x+y<1}dx dy \frac {y}{(\frac
{y}{\Delta}-(\frac {y}{2\Delta})^{2})^{1+\epsilon /2}\Delta ^{3+
e }} \nonumber \\
&&= {\int _{0}^{1}\int _{0}^{1}}ds dt \frac {s \times
st}{(st)^{1+\epsilon /2} \Delta (\Delta -\frac {st}{4})^{1+\epsilon /2}}
\nonumber  \\
&&=\int _{0}^{1}\int _{0}^{1}ds dt \frac {1}{s^{1+\epsilon }t^{\epsilon /2}
(1-\frac {s}{4})[1-\frac {s}{4}-\frac {t}{4}]^{1+\epsilon /2}} .
\label{eq:3l22a}
\end{eqnarray}

The $s$ dependent part in Eq.~(\ref{eq:3l22a}) can be separated into:
\begin{eqnarray}
\lefteqn{\int _{0}^{1}ds\frac {1}{s^{1+\epsilon }(1-\frac {s}{4-t})
^{1+\epsilon /2}}
[\frac {1}{1-\frac {s}{4}}-1]+\int _{0}^{1}ds \frac {1}{s^{1+\epsilon
}(1-\frac {s}{4-t})^{1+\epsilon /2}} }\nonumber  \\
&=& \int _{0}^{1}ds \frac {\frac {1}{4}}{(1-\frac {s}{4-t})(1-
\frac {s}{4})}+\int _{0}^{1}ds \frac {\frac {1}{4-t}}{s^{\epsilon}
(1-\frac {s}{4-t}) ^{1+\epsilon /2}} \nonumber \\
&&+ \int _{0}^{1}ds\frac {1}{s^{1+\epsilon} }+O(\epsilon ).
\label{eq:3l22b}
\end{eqnarray}
 The first term in Eq.~(\ref{eq:3l22b}) can be integrated out as:
\begin{equation}
 \int _{0}^{1}ds \frac {\frac {1}{4}}{(1-\frac {s}{4-t})(1-
\frac {s}{4})}=[\ln (1-\frac {t}{4})-\ln (1-\frac {t}{3})].
\end{equation}
 Inserting it into Eq.~(\ref{eq:3l22a}), we inherit:
\begin{equation}
\int _{0}^{1}dt (\frac {4-t}{t})[\ln (1-\frac {t}{4})-\ln
(1-\frac
{t}{3})]=4(I_D-I_E)+2\ln (\frac {3}{2})-3\ln (\frac {4}{3}).
\end{equation}
Again, inserting the second term in 
Eq.~(\ref{eq:3l22b}) into Eq.~(\ref{eq:3l22a}),
 one has:
\begin{equation}
\int _{0}^{1}dt \frac {1}{1-\frac {t}{4}}[-\ln (1-\frac {1}{4-
t})] = I_A-I_B+4\ln ^2(\frac {4}{3}).
\end{equation}

After substituting the 3nd term in Eq.~(\ref{eq:3l22b}) into
Eq.~(\ref{eq:3l22a}), one obtains:
\begin{eqnarray}
\lefteqn{\int _{0}^{1}ds \frac {1}{s^{1+\epsilon }}\cdot\int _{0}^{1}dt
[\frac {1}{1-\frac {t}{4}}-\frac {\epsilon }{2}\frac {[\ln (t)+\ln
(1-\frac {t}{4})]}{1-\frac {t}{4}}] }\nonumber  \\
&=&(\frac {1}{-\epsilon })[4\ln (\frac {4}{3})+\frac {\epsilon }
{2}(-I_C-I_A)] .
\end{eqnarray}

Here we  evaluate the integral of $3.2-3$.
\begin{eqnarray}
\lefteqn{3.2-3} \nonumber \\
&=&\frac {1}{4}\int _{-\infty}^{\infty} d^d \vec p\int _{-
\infty}^{\infty} d^d \vec q\int _{-\infty}^{\infty} d^d \vec k
\frac {\vec p\cdot\vec q }{(p^2+m^2)(q^2+m^2)}\int _{0}^{1}dx
\nonumber  \\
&& \frac {1}{(k^2-\vec p\cdot\vec k+p^2+xq^2-x \vec p\cdot\vec q
+m^2)^{2}}\nonumber  \\ &=&
\frac {1}{4}\int _{-\infty}^{\infty} d^d \vec p
\int _{-\infty}^{\infty} d^d \vec q\frac {\vec p\cdot\vec q}
{(p^2+m^2)(q^2+m^2)}
\int _{0}^{1}dx \pi ^{d/2}\frac {\Gamma (1-\epsilon /2)}{\Gamma (2)}
\nonumber \\
&&\frac {1}{(\frac {3}{4}p^2-x \vec p\cdot\vec q +xq^2
+m^2)^{1-\epsilon /2}}\nonumber  \\
&=& \frac {1}{4}\quad \pi ^{d/2}\frac {\Gamma (1-\epsilon/2)}
{(\frac {3}{4})^{1-\epsilon /2}}
\int _{0}^{1}dx
\int _{-\infty}^{\infty} d^d \vec p \int _{-\infty}^{\infty} d^d
\vec q \frac {\vec p\cdot\vec q }{(p^2+m^2)(q^2+m^2)}
\nonumber  \\
&&\frac {1}{(\frac {3}{4})^{1-\epsilon /2}(p^2-\frac {4}{3}x\vec
p\cdot\vec q
+\frac {4}{3}q^2 +\frac {4}{3}m^2)^{1-\epsilon /2}}\nonumber  \\
&=& \frac {1}{4}\pi ^{d/2}\frac {\Gamma (2-\epsilon /2)}{(\frac
{3}{4})^{1-\epsilon /2}}\int _{0}^{1}\int _{0}^{1}dxdy \frac {\vec
p\cdot\vec q }{(q^2+m^2)}\frac {y^{-\epsilon /2}}{(p^2-\frac {4}{3}xy
\vec p\cdot\vec q+\frac {4}{3}xy q^2+\frac {4}{3}ym^2)^{2-\epsilon /2}}
\nonumber  \\&=&
\frac {1}{4}\pi ^{d/2}\frac {\Gamma (2-\epsilon /2)}{(\frac {3}{4})^{1-
\epsilon /2}} (\pi)^{d/2}\frac {\Gamma (1-\epsilon )}
{\Gamma (2-\epsilon /2)}  \int
_{0}^{1}\int _{0}^{1}dxdy \int _{-\infty}^{\infty} d^d \vec q
\frac {1}{(q^2+m^2)} \nonumber  \\&&
\frac {y^{-\epsilon /2}\quad
\vec q\cdot(\frac {2}{3}xy\vec  q)}
{[(\frac {4}{3}xy-(\frac {2}{3}xy)^2)q^2+\frac {4}{3}ym^2]
^{1-\epsilon }}\nonumber  \\&=&
\frac {1}{4}\frac {(\pi)^{d} \Gamma (1-\epsilon )}{(\frac {3}{4}
)^{1-\epsilon /2}}\int _{-\infty}^{\infty} d^d \vec q
\frac {q^2}{(q^2+m^2)}\frac {(\frac {2}{3}xy)y^{-\epsilon /2}}
{\Box ^{1-\epsilon }(q^2+\frac {4y}{3\Box}m^2)^{1-\epsilon }} 
\nonumber \\
&=& \frac {1}{4} \frac {(\pi)^d \Gamma (1-\epsilon)}{(\frac {3}{4})
^{1-\epsilon /2}}
\frac {2}{3} \frac {(\pi)^{d/2}\Gamma(2+\epsilon /2)\Gamma(-\frac 
{3}{2}\epsilon)}
{\Gamma(1-\epsilon)\Gamma(1+\epsilon /2)}(m^2)^{\frac {3}{2}\epsilon}
\nonumber \\
&&\int _{0}^{1} \int _{0}^{1}dx dy \frac 
{y^{-\epsilon /2}xy}{(\Box)^{1-\epsilon}}
(\frac {4y}{3\Box})^{\frac {3}{2}\epsilon} .
\label{eq:3l23i}
\end{eqnarray}
Here $\Box = \frac {4}{3}xy -(\frac {2}{3}xy)^2$.
The integration over $x$ and $y$ in Eq.~(\ref{eq:3l23i})    is obtained as:
\begin{eqnarray}
\lefteqn{\int _{0}^{1}\int _{0}^{1}dxdy (\frac {4}{3})^{3\epsilon   /2}\frac
{xy^{1+\epsilon }}{[\frac {4}{3}(xy)-\frac {4}{9}(xy)^{2}]
^{1+\epsilon /2}}}\nonumber  \\
&=&(\frac {4}{3})^{3\epsilon    /2}\int _{0}^{1}\int _{0}^{1}dx dy
\frac {xy^{1+\epsilon }}{(xy)^{1+\epsilon /2}(\frac {4}{3})^{1+\epsilon /2}}
\frac {1}{(1-\frac {xy}{3})^{1+\epsilon /2}}\nonumber  \\
&=&-3\int _{0}^{1} dy [\frac {1-\frac {y}{3}}{y}]  \nonumber \\
&=&-3I_E +O(\epsilon )  .
\end{eqnarray}

Finally, we summarize the results of two 3-loop calculations.
One  should observe that there are not any
sub-divergences in 3-loop diagrams and therefore
no such a term as $\ln (cm^2a^2)^2$  exists.
In the following summation of  both  3-loop calculations
results, we  demonstrate  this observation  by explicit
calculations.
By collecting all previous results,
 the contributions  of FD.3l1 and FD.3l2 in $\epsilon ^{-2}$  are given by:
\begin{eqnarray} 
\mbox{FD.3l1}&:&-\frac {1}{2}[\frac {1}{4}\Gamma 
(\frac {-3 \epsilon}{2})(-\frac
{8}{\epsilon}) \ln (\frac {4}{3})
-2 \frac {1}{8}\Gamma (\frac {-3\epsilon   }{2})(-\frac {4}{\epsilon}
)\ln (\frac {4}{3})-   \nonumber \\
&&2 \frac {1}{8}\Gamma (\frac {-3 \epsilon}{2})
4 \ln (\frac {4}{3}) \frac {-1}{\epsilon}]=0     \\
\mbox{FD.3l2}&:&
-\frac {1}{4}\Gamma (\frac {-3 \epsilon}{2})
4 \ln (\frac {4}{3})(-\frac {1}{\epsilon})
+2 (\frac {1}{8})\Gamma (\frac {-3 \epsilon}{2})
4 \ln (\frac {4}{3})(-\frac {1}{\epsilon}) =0
\end{eqnarray} 

These confirm our observation mentioned above.
The consequent results of the leading divergences  
which contributes to  $Z_g$  are used in Eq. ~\ref{eq:ggz}.

\section{Basic Integrals}
\label{a61}

We define $\Xi(x,s)=\sum _{n=1}^{\infty}\frac {x^n}{n^s}$,
and $\Phi (x,s)=\sum _{n=1}^{\infty}(-1)^{n}\frac {x^n}{n^s}$.

$I_A$. $\int _{0}^{1}du \frac {\ln (1-\frac {u}{4})}{(1-\frac
{u}{4})}=-2 \ln ^2(\frac {3}{4})$

$I_B$. $\int _{0}^{1}du \frac {\ln (1-\frac {u}{3})}{(1-\frac
{u}{4})}=2\ln ^2(\frac {4}{3})-4[\Xi(\frac {1}{3},2)-
\Xi (\frac {1}{4},2)]$

$I_C$. $\int _{0}^{1}du \frac {\ln u}{(1-\frac {u}{4})}=
-4\Xi (\frac {1}{4},2) $

$I_D$. $\int _{0}^{1}du \frac {\ln (1-\frac {u}{4})}{u}
=-\Xi (\frac {1}{4},2)$

$I_E$. $\int _{0}^{1}du \frac {\ln (1-\frac {u}{3})}{u}
=-\Xi (\frac {1}{3},2)$

$I_F$. $\int _{0}^{1}\int _{0}^{1}dudv \frac {\ln (1-\frac
{uv}{4})}{u}=-[\Xi (\frac {1}{4},2)+3\ln (\frac {4}{3})-1]$

$I_G$. $\int _{0}^{1}du  \ln (u)\ln (1-\frac {u}{4})
=2-3\ln (\frac {4}{3})-4 \Xi (\frac {1}{4},2)$

$I_H$. $\int _{0}^{1}\int _{0}^{1}du dv \frac {\ln (1-\frac
{u}{4-
v})}{u}=4 \ln (\frac {4}{3})-\frac {5}{2}\ln (\frac {3}{2})+\frac
{1}{2}$

$I_J$. $\int _{0}^{1} du \ln (1-\frac {u}{4})
=3\ln (\frac {4}{3})-1$

\newpage

\begin{center}
{\bf  Figures Caption}
\end{center}

\vskip 0.1in
\underline{Fig. 1}

A two-dimensional cut (along a lattice plane perpendicular to the disordered
substrate) of the three-dimensional system.

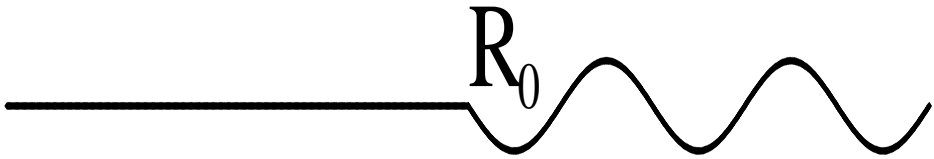
\underline{Fig. 2}

The Feynnman diagram for the correlation function and response function.

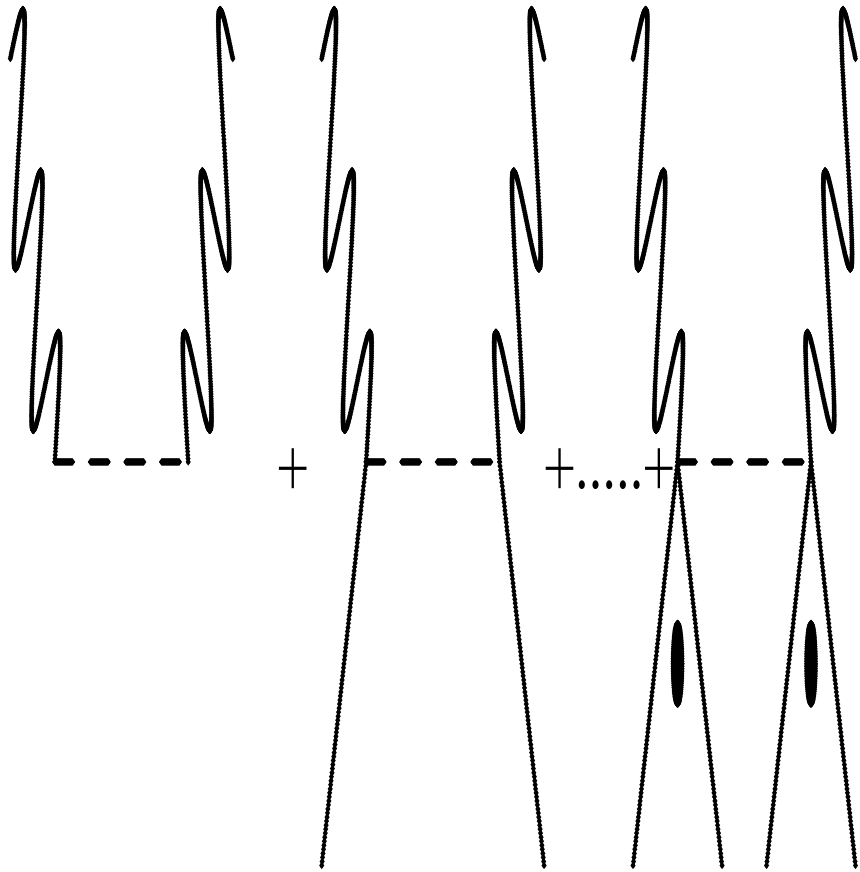
\underline{Fig. 3}

The Feynman diagram representing $\tilde{\phi} 
(\vec x,t) \tilde{\phi}(\vec x,
t') \cos (\phi(\vec x ,t)- \phi(\vec x, t'))$.

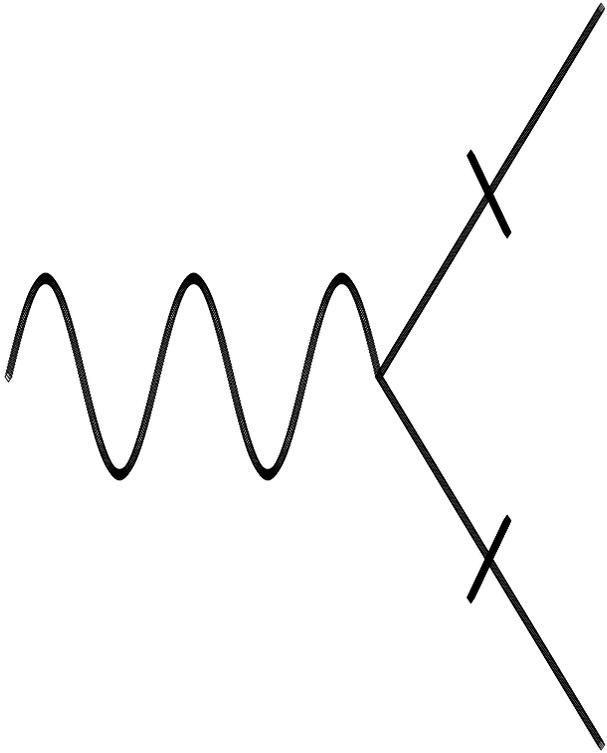
\underline{Fig. 4}

The basic diagram for $\lambda $.

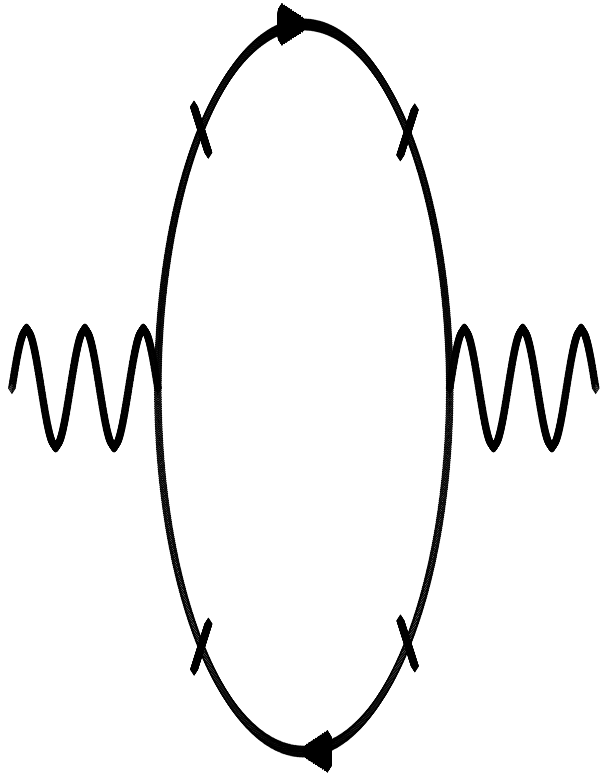
\underline {Fig. 5}

The  Feynman diagram   for $\Gamma _{2,0}$ up to order $\lambda  ^2$.

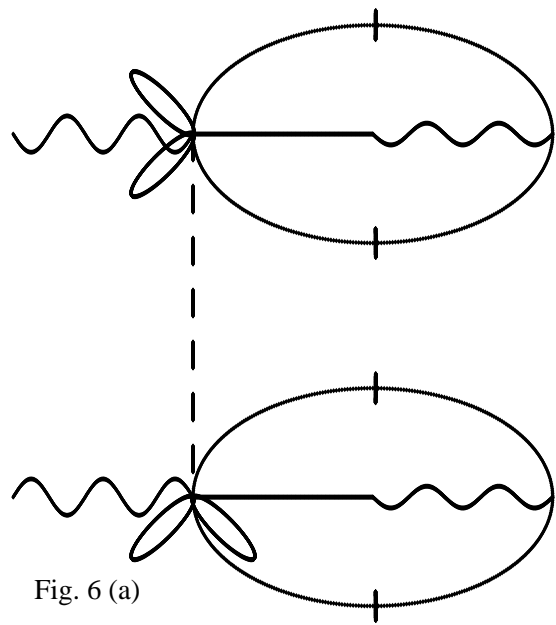
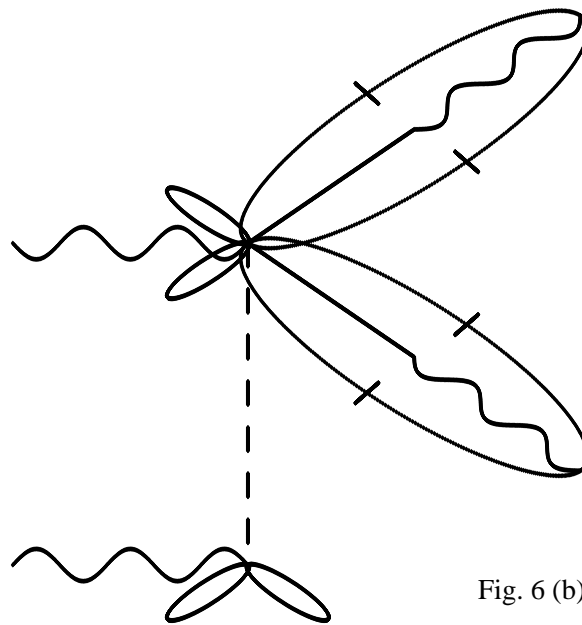
\underline {Fig. 6}

Two mutually canceled  two-loop Feynman diagrams.

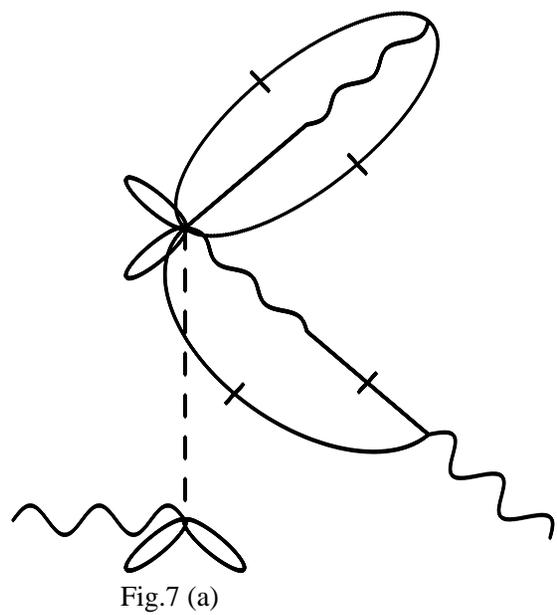
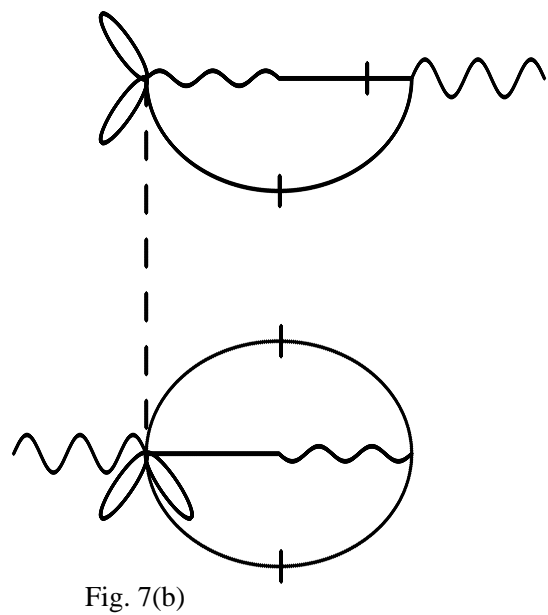
\underline{Fig. 7}
Two mutually canceled  two-loop Feynman diagrams.

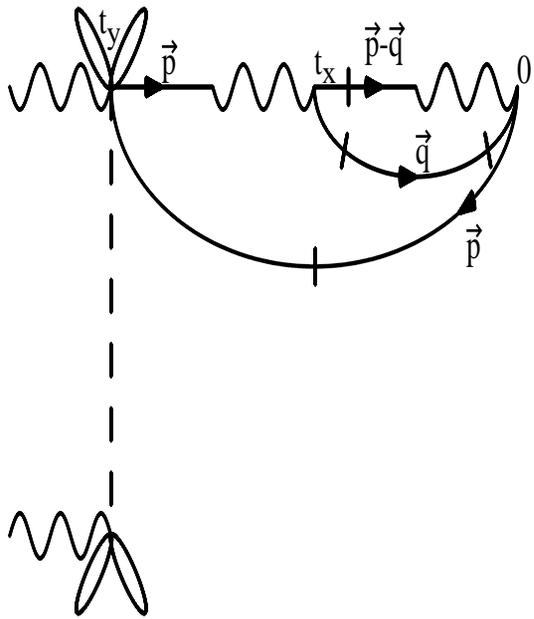
\underline{Fig. 8}

A 2-loop Feynman  diagram: FD. 2l1

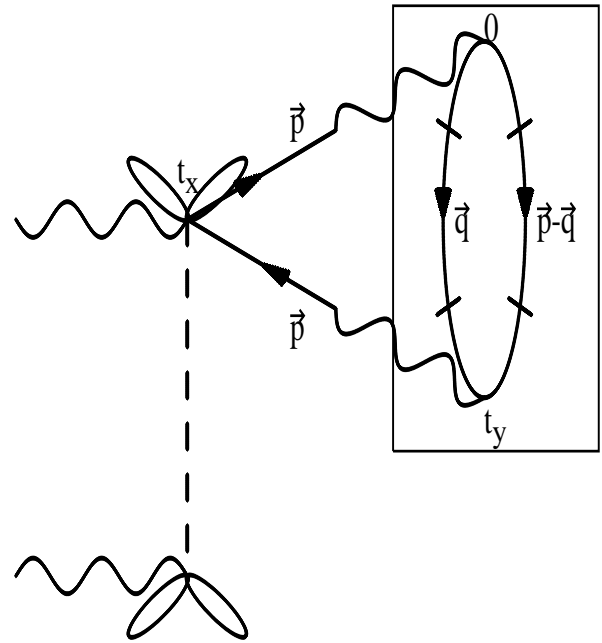
\underline{Fig. 9}

A 2-loop Feynman  diagram: FD. 2l2

\underline{Fig. 10}
A 2-loop Feynman  diagram: FD. 2l3

\underline{Fig. 11}

A 2-loop Feynman  diagram: FD. 2l4

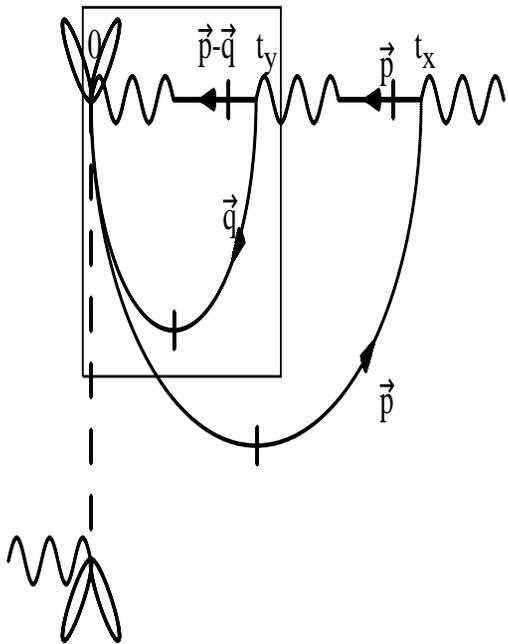
\underline{Fig. 12}

A 2-loop Feynman  diagram: FD. 2l5

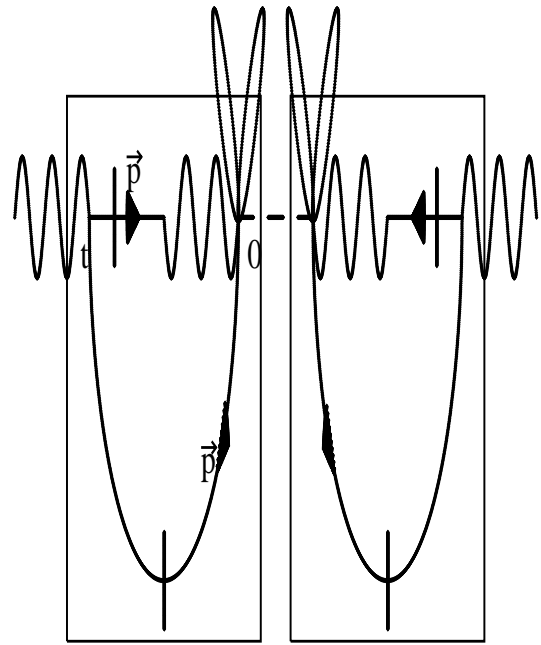
\underline{Fig. 13}

A 2-loop Feynman  diagram: FD. 2l6

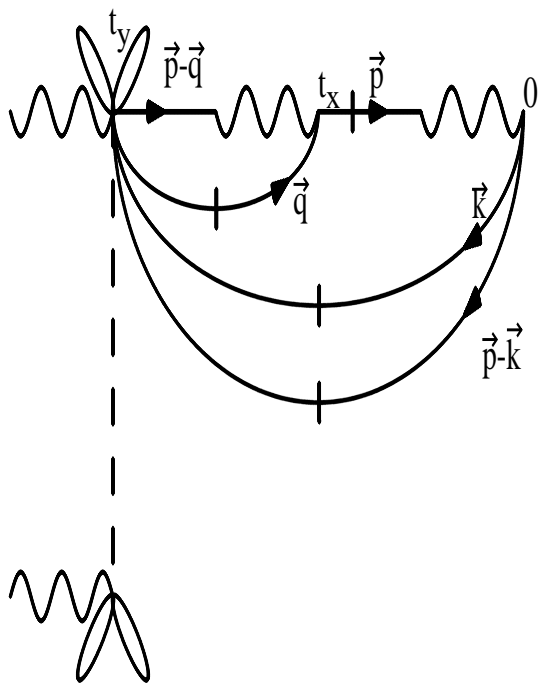
\underline{Fig. 14}

A 3-loop Feynman  diagram: FD. 3l1

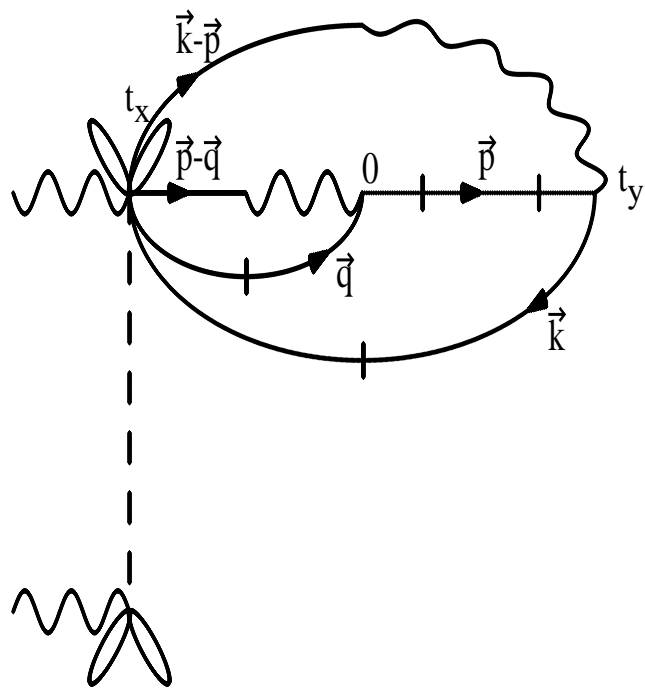
\underline{Fig. 15}

A 3-loop Feynman  diagram: FD. 3l2

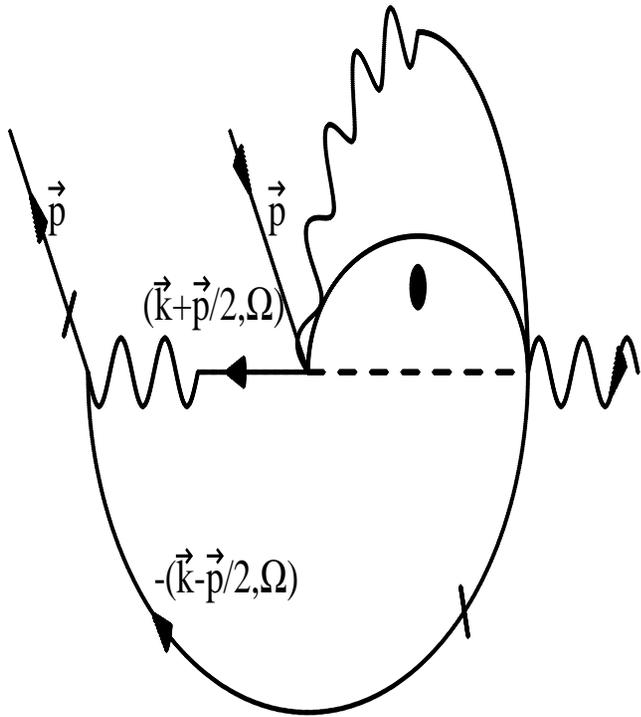
\underline{Fig. 16}

A  Feynman  diagram: FD. nla

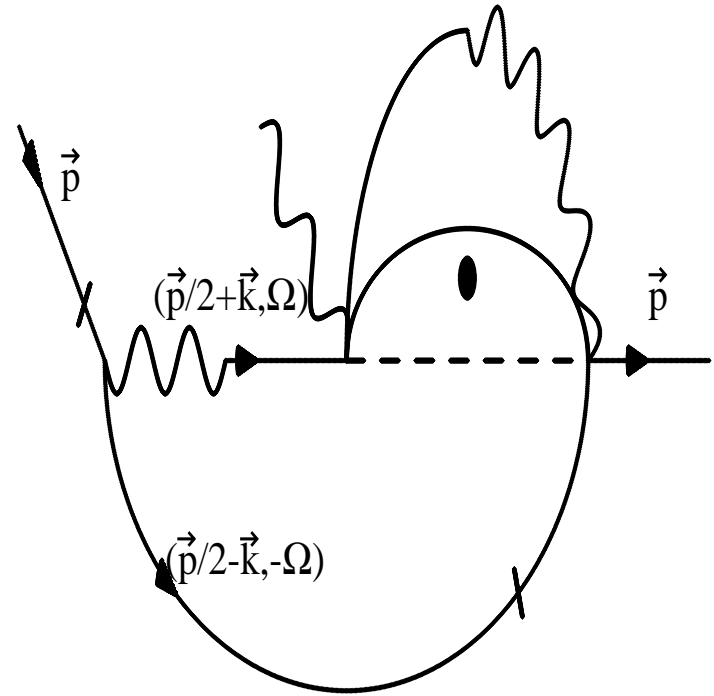
\underline{Fig. 17}

A  Feynman  diagram: FD. nlb

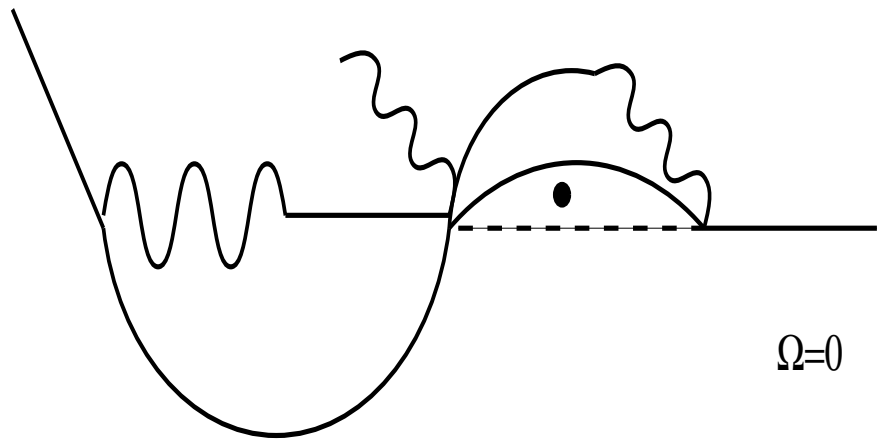
\underline{Fig. 18}

Two  Feynman  diagrams contributing to $\Gamma_{1,2}$.

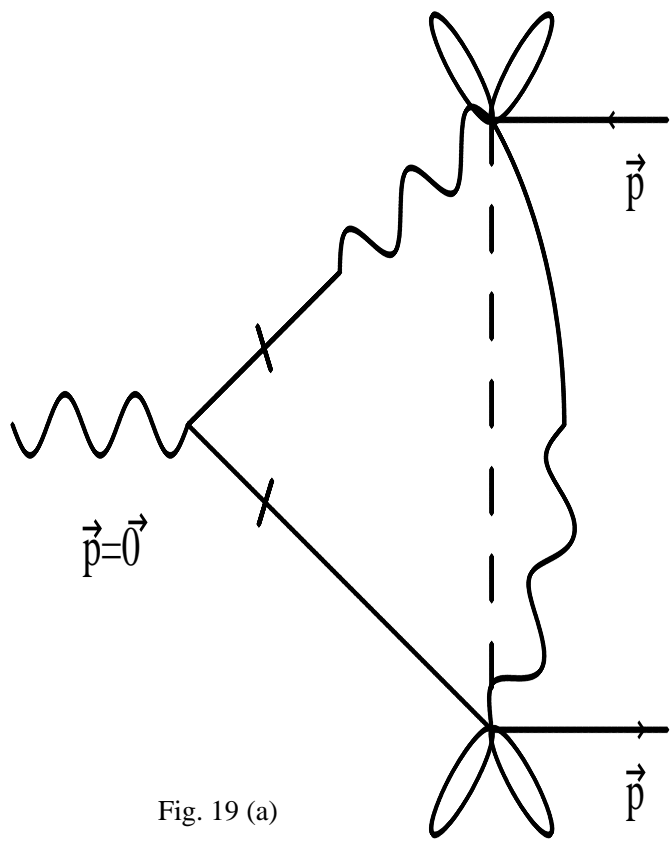
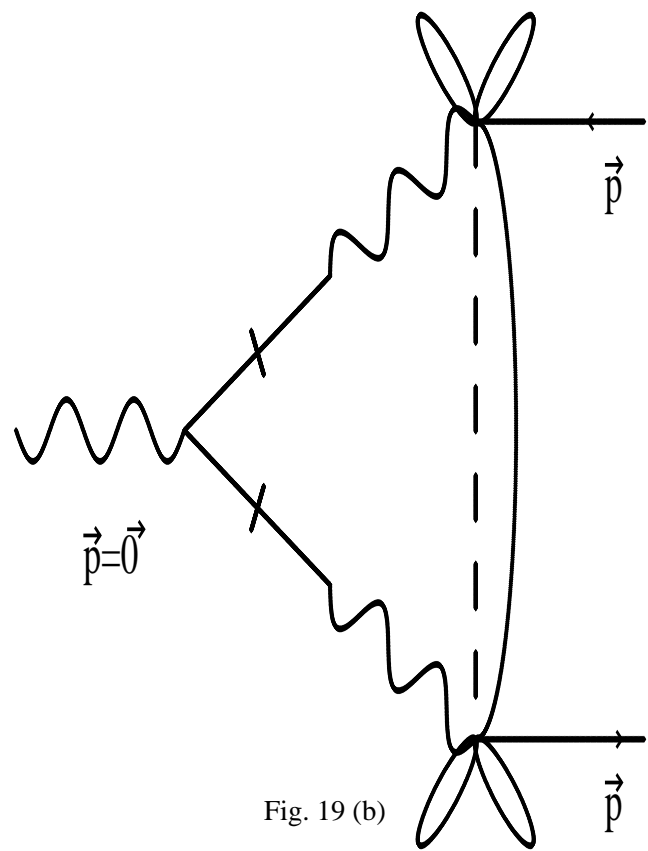
\underline{Fig. 19}

The other two Feynman  diagrams contributing to $\Gamma_{1,2}$.
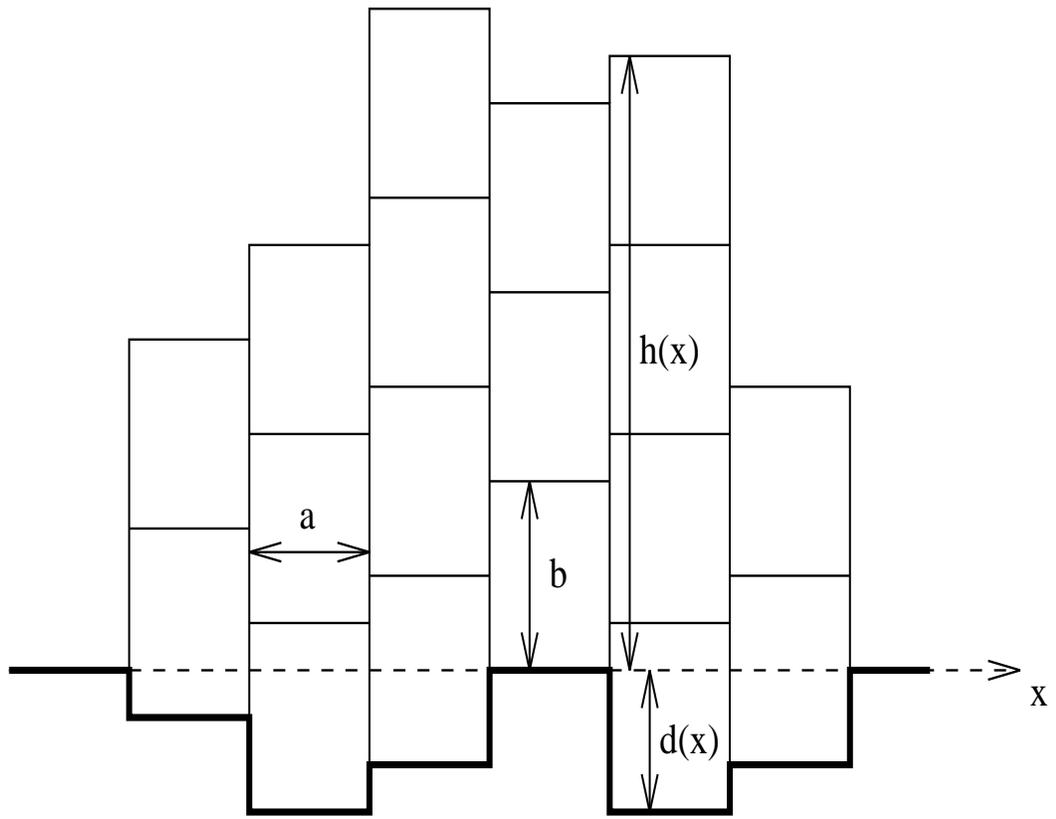
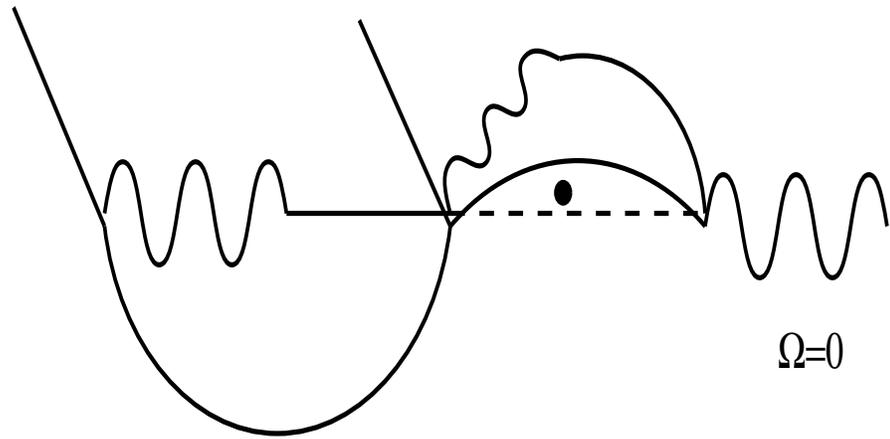
\end{document}